\documentclass[13pt]{iopart}

\usepackage{iopams}
\expandafter\let\csname equation*\endcsname\relax
\expandafter\let\csname endequation*\endcsname\relax
\usepackage{amsmath}
\usepackage{graphicx}
\usepackage{xcolor}
\begin{document}

\title[$\beta$-decay of N=126 isotones for the  $r$-process nucleosynthesis]{$\beta$-decay of N=126 isotones for the $r$-process nucleosynthesis}
\author{Jameel-Un Nabi$^{1,2}$, Necla \c{C}akmak$^{3}$, Asim Ullah$^{2}$\footnote{corresponding author email: asimullah844@gmail.com } and Asad Ullah Khan$^{2}$}

\address{$^{1}$University of Wah, Quaid Avenue, Wah Cantt 47040, Punjab, Pakistan}
\address{$^{2}$Faculty of Engineering Sciences, GIK Institute of Engineering Sciences and Technology, Topi 23640, Khyber Pakhtunkhwa, Pakistan}
\address{$^{3}$Faculty of Science, Department of Physics, University of Karab\"uk, 78050, Karab\"uk, Turkey}

\vspace{10pt}
\begin{abstract}
{The $\beta$-decay properties of nuclei with neutron 
	number $N$ = 126 is investigated in this paper. Two different versions of the proton-neutron quasi particle random phase (pn-QRPA) model were employed to
	compute $\beta$-decay rates and half-lives for the $N$ = 126 isotones. The first set of calculation solves the pn-QRPA equations using the schematic model (SM) approach. The Woods-Saxon potential was employed as a mean-field basis. A spherical shape assigned for each waiting point nuclei throughout all simulations. Both allowed Gamow-Teller (GT) and first-forbidden (FF) transitions were considered in the particle-hole (\textit{ph}) channel. 
	The second set uses the pn-QRPA model  in deformed Nilsson
	basis to calculate $\beta$-decay rates for allowed GT and unique first-forbidden (U1F) transitions under
	terrestrial and stellar conditions. Our results are in agreement with shell model findings
	that first-forbidden transitions lead to a considerable decrement in the 
	calculated half-lives of the isotones. Inclusion  of the first-forbidden
	contribution led to a  decent agreement of our computed terrestrial $\beta$-decay half-lives with measured ones, much better than the previous calculations. The possible implication of the waiting point nuclei on $r$-process nucleosynthesis is discussed briefly.}

\end{abstract}
%
\vspace{2pc}
\noindent{\rm Keywords}: Gamow-Teller transitions, first-forbidden transitions, pn-QRPA theory, waiting-point nuclei, $\beta$-decay rates, $r$-process
\ioptwocol
\maketitle
\section{Introduction:}
Quest for a better understanding of the $r$-process continues to date. The interest is primarily due to the fact that nearly half of the heavy elements beyond iron is thought to be
synthesized during the $r$-process \cite{Bur57, Cow91}. At closed-shell ($N$ =
50, 82, 126), the $r$-process flow of matter decelerates. The
corresponding nuclei have to wait for several $\beta$-decays to
occur before capturing of neutrons resumes (also referred to as
waiting points). The matter is therefore accumulated at these waiting
points resulting in the well-known peaks in the observed $r$-process
abundance distribution. The challenge posed to theorists is to come
up with a microscopic $r$-process nucleosynthesis calculation
reproducing this observed distribution pattern.
The $\beta$-decay rate of waiting points is one of the key nuclear properties that can affect
the $r$-process matter flow.The $\beta$-decay rates are
important not only in the accurate determination of the structure of the stellar core but also play a vital role in the elemental abundance and nucleosynthesis calculations.
A reliable calculation of $\beta$-decay
half-lives for the waiting points is one of the pre-requisites for a  better understanding of the $r$-process and
reproduction of the observed abundance curve of the $r$-process
nuclei. The experimental data is rather scarce but is expected to
improve with results from new heavy-ion accelerator facility
(e.g. RIKEN, GSI-FRS, GANIL-LISE and CERN-ISOLDE). Despite advances in measurements of $\beta$-decay half-lives, a better understanding of the $r$-process
can be realized with reliable theoretical estimates of $\beta$-decay
properties of waiting points specially under $r$-process physical
conditions ($T \sim$ 10$^{9}$\textit{K}, neutron densities $>$ 10$^{20}$
\textit{cm$^{-3}$}). The observed $r$-process spectra of waiting-point nuclei may be affected by the
presence of low-lying energy levels possessing different parities. This necessitates
the incorporation of the first-forbidden (FF) chapter to the $\beta$-decay
half-lives. 
For the $N$ = 50 and 82 isotones, the past theoretical calculations
of $\beta$-decays are in decent comparison with each other as against those
for $N$ = 126 isotones \cite{Suz12}. One important reason for
disagreement between different theoretical calculations is the
computation of FF transitions. Single-particle states appearing with unlike parity can significantly affect the calculated half-lives. The
FF transitions, therefore, become important for the $N$ = 126 isotones
in addition to the allowed Gamow-Teller (GT) transitions.
Few noticeable calculations of $\beta$-decay half-lives
of $N$ = 126 isotones include (i) quasiparticle random phase
approximation (QRPA) calculation of GT \cite{Mol97} and gross theory calculation
of GT and FF rates \cite{Mol03} (referred to as QRPA-FRDM), (ii) the
continuum QRPA approach based on the self-consistent ground state
description in the framework of the density functional theory
\cite{Bor06,Bor11} (referred to as CQRPA-DF3), (iii) shell model rates
\cite{Suz12} (referred to as SM), (iv) large-scale shell model
rates \cite{Zhi13} (referred to as LSSM) (v) an empirical formula based calculation \cite{Zhou17} and calculations of QRPA within energy-density functional theory ~\cite{Mark16, Engel99, Ney20}.
In this work, two different sets of the pn-QRPA model were employed to study $N$ = 126
isotones. The first set of calculation solves the pn-QRPA equations
using the spherical schematic model approach. The Woods-Saxon (WS)
potential was employed as mean-field basis and deformation of
waiting points were taken as zero. Allowed GT and FF (rank 0, 1 and 2) transitions were
calculated separately for computing $\beta$-decay
half-lives. The second set uses a pn-QRPA model  in deformed Nilsson
basis to calculate $\beta$-decay rates (allowed GT and unique first-forbidden (U1F)) under
terrestrial and stellar conditions. The first and second sets of the pn-QRPA model are represented as pn-QRPA (WS) and pn-QRPA (N), respectively, throughout this manuscript. 
The paper is divided into four broad sections. Section~2 outlines the brief formalism involved in our calculation. 
Results and comparison with previous theoretical and measured data
(wherever available) are presented in Section~3. Conclusions are finally stated in Section~4.
\section{Formalism}
A brief formalism of both versions of the pn-QRPA model is presented here.

\subsection{The pn-QRPA (WS) model}
Allowed GT and the FF $\beta$-decay half-lives are calculated employing spherical schematic model (SSM) within the pn-QRPA framework. 
The Woods-Saxon potential was used as a mean-field basis. The eigenvalues and eigenfunctions of the Hamiltonian with separable residual GT and FF effective interaction were only calculated in particle-hole (\textit{ph}) channel. We shall consider the GT $1^{+}$ excitations in odd-odd nuclei generated from the correlated ground state of the parent nucleus by the charge-exchange spin-spin forces and use the eigenstates of the single quasiparticle Hamiltonian $H_{sqp}$ as a basis. The schematic method Hamiltonian for GT excitations in the neighboring odd-odd nuclei is given in the following form:
\begin{eqnarray}
	H_{SSM}=H_{sqp}+h_{ph},
\end{eqnarray}
where $H_{sqp}$ is the single quasiparticle (\textit{sqp}) Hamiltonian and $h_{ph}$ is the GT effective interaction in the \textit{ph} channel. Details of solution of allowed GT formalism are available in \cite{Cak10a,Cak12}.

A separable FF force with the \textit{ph} channel was employed aiming to reduce the eigenvalue equation to an easily solvable algebraic equation of fourth order and minimize the computational effort. The spherical schematic method Hamiltonian for FF excitations is given by
\begin{eqnarray}
	H_{SSM}=H_{sqp}+h_{ph}.
\end{eqnarray}
The spin-isospin effective interaction denoted by $\hat{h}_{ph}$ causes $0^{-}$, $1^{-}$, $2^{-}$ vibrational modes in \textit{ph} channel, and is specified as
\begin{eqnarray}
	\hat{h}_{ph}={2\chi_{ph}}\sum_{j_{p}j_{n}j_{p'}j_{n'}}[b_{j_{p}j_{n}}A^{\dag}_{j_{p}j_{n}}+\bar{b}_{j_{p}j_{n}}A_{j_{p}j_{n}}] \nonumber \\ ~~~~~~~~~~~~~~~~~~~\times[b_{j_{p'}j_{n'}}A_{j_{p'}j_{n'}}+\bar{b}_{j_{p'}j_{n'}}A^{\dag}_{j_{p'}j_{n'}}],
\end{eqnarray}
where $\chi_{ph}$ stands for the \textit{ph} effective interaction constant, which was taken as $\chi_{GT}=5.2A^{0.7} MeV$, $\chi_{rank 0}=30A^{-5/3} MeVfm^{-2}$, $\chi_{rank 1}=99A^{-5/3} MeVfm^{-2}$ and $\chi_{rank 2}=350A^{-5/3} MeVfm^{-2}$ for allowed GT, rank 0, rank 1 and rank 2 transitions, respectively. The effective interaction constant in the \textit{ph} channel was fixed from the experimental value of the resonance energy.

The operator $A^{\dag}_{j_{p}j_{n}}$ ($A_{j_{p}j_{n}}$) is called the quasi-boson creation (annihilation) operator and specified as follows
\begin{eqnarray}
	A^{\dag}_{j_{p}j_{n}}=\frac{1}{\sqrt{2j+1}}\sum_{m}(-1)^{j-m}\alpha^{\dag}_{j_{p}m_{n}}\alpha_{j_{p}-m_{n}},
\end{eqnarray}
\begin{eqnarray}
	A_{j_{p}j_{n}}=(A^{\dag}_{j_{p}j_{n}})^{\dag}
\end{eqnarray}
where $\alpha^{\dag}_{j_{p}m_{n}}$ presents the quasiparticle creation (annihilation) operator. The $\bar{b}_{j_{p}j_{n}}$, ${b}_{j_{p}j_{n}}$ appearing above stands for reduced matrix elements of the multipole operators for rank 0, 1 and 2 \cite{Boh69}. Here we have shown only the relevant equations and detailed formalism can be seen from \cite{Nab16, Nab17}. Details of rank 0 excitation computation are available in \cite{Cak10}. The nuclear matrix elements (both the relativistic and the non-relativistic, respectively) for $\lambda^{\pi}=0^{-}$ are given by
\begin{eqnarray}
	M^{\pm}(\lambda=0, \rho_{A})=\frac{g_{A}}{{(4\pi)}^{1/2}c}\Sigma_{k}t_{\pm}(k)(\vec{\sigma}_{k}\cdot\vec\vartheta_{k}),
\end{eqnarray}
\begin{eqnarray}
	M^{\pm}(\lambda=0, \kappa=1, j_{A})=g_{A}\Sigma_{k}t_{\pm}(k)r_{k}(Y_{1}(r_{k})\sigma_{k})_{0}.
\end{eqnarray}
where $\sigma_k$ stands for the Pauli-spin matrices and $t_{\pm}$ represents the iso-spin raising/lowering operator. 
Similarly, the relativistic and the non-relativistic matrix elements, for $\lambda^{\pi}=1^{-}$ are presented by
\begin{eqnarray}
	M^{\pm}(\lambda=1,\kappa=0,\mu,j_{v})= \nonumber
\end{eqnarray}
\begin{eqnarray}
~~~~~~~~   \frac{g_{v}}{{(4\pi)}^{1/2}c}\Sigma_{k}t_{\pm}(k)r_{k}(\vec\vartheta_{k})_{1\mu},
\end{eqnarray}
\begin{eqnarray}
	M^{\pm}(\lambda=1,\rho_{v},\mu)=g_{v}\Sigma_{k}t_{\pm}(k)r_{k}Y_{1\mu}(r_{k}),
\end{eqnarray}
\begin{eqnarray}
	M^{\pm}(\lambda=1,\kappa=1,j_{v},\mu)= \nonumber
\end{eqnarray}
\begin{eqnarray}
~~~~~~~~   g_{A}\Sigma_{k}t_{\pm}(k)r_{k}(Y_{1}(r_{k})\sigma_{k})_{1\mu}.
\end{eqnarray}
Finally, the non-relativistic matrix element for $\lambda^{\pi}=2^{-}$ is specified by
\begin{eqnarray}
	M^{\pm}(\lambda=2,\kappa=1,j_{A},\mu)= \nonumber
\end{eqnarray}
\begin{eqnarray}
~~~~~~~~   g_{A}\Sigma_{k}t_{\pm}(k)r_{k}(Y_{1}(r_{k})\sigma_{k})_{2\mu}.
\end{eqnarray}
The transitions probabilities $B(\lambda^{\pi}=0^{-},1^{-},2^{-}; \beta^{\pm})$ are specified by \cite{Boh69}
\begin{eqnarray}
	B(\lambda^{\pi}=0^{-} \beta^{\pm})=|<0^{-}_{i}\|M_{\beta^{\pm}}^{0}\|0^{+}>|^{2},
\end{eqnarray}
where
\begin{eqnarray}
	M_{\beta^{\pm}}^{0}=\pm M^{\pm}(\lambda=0, \rho_{A}) \nonumber
\end{eqnarray}
\begin{eqnarray}
~~~~~~~~   -i\frac{2{\pi}m_{e}c}{h}\xi M^{\pm}(\lambda=0,\kappa=1,j_{A}).
\end{eqnarray}
\begin{eqnarray}
	B(\lambda^{\pi}=1^{-}, \beta^{\pm}) = |<1^{-}_{i}\|M_{\beta^{\pm}}^{1}\|0^{+}>|^{2},
\end{eqnarray}
where
\begin{eqnarray}
	M_{\beta^{\pm}}^{1}=M^{\pm}(\lambda=1,\kappa=0,j_{v},\mu) \nonumber \\
~~~~~~~~	\pm	i\frac{2{\pi}m_{e}c}{{3}^{1/2}h} M^{\pm}(\lambda=1,\rho_{v},\mu)\nonumber
\end{eqnarray}
\begin{eqnarray}
~~~~~~~~	+ i{(\frac{2}{3})}^{1/2}\frac{2{\pi}m_{e}c}{h}\xi
	M^{\pm}(\lambda=1,\kappa=1,j_{A},\mu).
\end{eqnarray}
\begin{eqnarray}
	B(\lambda^{\pi}=2^{-},\beta^{\pm})=|<2^{-}_{i}\|M_{\beta^{\pm}}^{2}\|0^{+}>|^{2},
\end{eqnarray}
and
\begin{eqnarray}
	M_{\beta^{\pm}}^{2}= M^{\pm}(\lambda=2,\kappa=1,j_{A},\mu).
\end{eqnarray}
In Eqs.(12, 14), the upper sign stands for $\beta^{+}$ while the lower signs show the $\beta^{-}$ decay. The \textit{ft} values are specified by
\begin{eqnarray}
	(ft)_{\beta^{\pm}}=\frac{D}{(g_{A}/g_{V})^{2}4\pi B(I_{i}\longrightarrow I_{f},\beta^{\pm})}.
\end{eqnarray}
Transitions with $\lambda=n+1$ are known as U1F transitions \cite{Boh69} and the $ft$ values are calculated using
\begin{eqnarray}
	(ft)_{\beta^{\pm}}=\frac{D}{(g_{A}/g_{V})^{2}4\pi B(I_{i}\longrightarrow I_{f},\beta^{\pm})} \nonumber
\end{eqnarray}
\begin{eqnarray}
~~~~~~~~~  \times\frac{(2n+1)!!}{[(n+1)!]^{2}n!},
\end{eqnarray}
where $D=6295\textit{sec.}$ and the effective ratio of axial and vector coupling constants is taken as $(g_{A}/g_{V})=-1,254$. 
No explicit quenching factor was introduced in our calculation. The pair correlation function was chosen as $C_{n} = C_{p} = 12 / \sqrt{A}$ for the open shell nuclei. The energies were calculated from ground state of the daughter nuclei in all calculations. To obtain the beta transition probabilities, the nuclear matrix elements were calculated as a whole with the relativistic and the non-relativistic terms for the rank 0 and rank 1 transitions. Thus, FF contributions to the nuclear matrix element were computed by considering the relativistic correction terms. Hence, the contributions coming from the virtual $0^{-}$ and $1^{-}$ intermediate states were obtained within $\xi-$ approximation. As can be seen from Eqs. 13 and 15, the $\xi-$ approximation is only considered in the calculation of the non-relativistic terms in the rank 0 and rank 1 excitations. According to the Bohr-Mottelson model, the $\xi-$ approximation was taken into account in the investigation of the first forbidden transitions. This approximation is fairly accurate for the investigated transitions, but it is important in the quantitative evaluation of the multipole moments to include the corrections due to the finite nuclear size. In the present paper, this approximation was applied for the first time in the N=126 nuclei. The unique first forbidden $2^{-}$ contributions do not contain any relativistic term. The FF transitions become significant for neutron-rich isotopes. Therefore, the contribution of the FF excitations is very important when calculating the total $\beta$-decay half-lives.

\subsubsection{Extension for odd-A Nuclei}

In this section, we present a brief summary of the necessary formalism for the odd-A nuclei. The wave function of the odd mass (with odd neutron) nuclei  in the pn-QRPA(WS) method is given by
\begin{eqnarray}
	\arrowvert\Psi^{j}_{j_{k}m_{k}}>=\Omega^{j\dag}_{j_{k}m_{k}}\arrowvert 0>= \nonumber \\
	N^{j}_{j_{k}}\alpha^{\dag}_{j_{k}m_{k}}+\sum_{j_{\nu}m_{\nu}}R^{ij}_{k\nu}A^{\dag}_{i}\alpha^{\dag}_{j_{\nu}m_{\nu}}\arrowvert 0>
\end{eqnarray}
where $\Omega^{j\dag}_{j_{k}m_{k}}$ and $\arrowvert 0>$ present the phonon operator, the phonon vacumm, respectively.  Also, $N^{j}_{j_{k}}$ and $R^{ij}_{k\nu}$ are the quasiboson amplitutes corresponding to the states and are fulfilled by the normalization conditions. It is that the wave function is formed by superposition of one and three quasiparticle (one quasiparticle $+$ one phonon) states. The equation of motion the pn-QRPA (WS) is given by 
\begin{eqnarray}
	[H_{SSM},\Omega^{j\dag}_{j_{k}m_{k}}] \arrowvert 0>=\omega^{j}_{j_{k}m_{k}}\Omega^{j\dag}_{j_{k}m_{k}}\arrowvert 0>
\end{eqnarray}
The excitation energies $\omega^{j}_{j_{k}m_{k}}$ and the wave functions of the GT and FF excitations are obtained from the pn-QRPA(WS) equation of motion. 
One-particle and one-hole nuclei allow of the simplest possible theoretical description of their states. The structure of one-particle nuclei within the simple mean-field approximation is the following. One-proton states $\arrowvert \nu>$ and one-neutron states $\arrowvert k>$ are described as
\begin{eqnarray}
	\arrowvert \nu>=A^{\dag}_{\nu}|core>,~~~\arrowvert k>=A^{\dag}_{k}|core> \nonumber
\end{eqnarray}
where $|core>$ is the core with its Fermi level at some magic number. The details of the solution of the GT and FF transitions for the odd$-$A nuclei can be seen from \cite{Cak10b, Sel19}.

\subsection{The pn-QRPA (N) model}
Our second set of calculation involves solution of pn-QRPA equations in the deformed Nilsson basis.
The Hamiltonian of the model is given by
\begin{equation}
	H^{QRPA} = H^{sp} + V^{pair} + V ^{ph}_{GT} + V^{pp}_{GT}.
	\label{Eqt. 23}
\end{equation}
where $H^{sp}$, $V^{pair}$, $V_{GT}^{ph}$ and $V_{GT}^{pp}$ represent single-particle Hamiltonian, pairing potential, particle-hole GT force and particle-particle GT force, respectively. Single particle wave functions and energies were computed in the
deformed Nilsson basis. Pairing was treated within the BCS
formalism. It is to be  noted that  both
particle-hole (\textit{ph}) and particle-particle (\textit{pp}) channels  were
considered in the GT force in the current model. The \textit{ph} interaction strength $\chi$ was taken as 4.2/A \textit{MeV} and 56.16/A \textit{MeV fm$^{-2}$} for the allowed GT and U1F transitions respectively, following 1/A dependence \cite{hom96}. The \textit{pp} interaction strength was taken as 0.0001. The functional form of $\chi$ is the same as previously used in \cite{Nab15,Nab16,Nab17}. The deformation parameter ($\beta_2$) was calculated using $\beta_2$ = $\frac{125 Q_2}{1.44ZA^{2/3}}$ where the quadrupole moment $Q_2$ was taken from Ref.~\cite{Mol81, Mol16}. Details of solution of the
Hamiltonian [Eq.~(19)] may be seen from \cite{Mut92}. Computation of
terrestrial $\beta$-decay half-lives may be seen from \cite{Sta90}.
Below we present a brief formalism involved in the calculation of stellar
weak rates using the current model.

The stellar $\beta$-decay rates from the $\mathit{i}$th parent state to the $\mathit{j}$th daughter state of the nucleus is given by

\begin{equation}\label{lij}
	\lambda_{ij}^{\beta} =
	\frac{m_{e}^{5}c^{4}}{2\pi^{3}\hbar^{7}}\sum_{\Delta
		J^{\pi}}g^{2}B_{ij}(\Delta J^{\pi})f_{ij}(\Delta J^{\pi}).
\end{equation}
In above equation $B_{ij}(\Delta J^{\pi})$  stands for $\beta$-decay reduced transition probability while $ f_{ij}(\Delta J^{\pi})$ is the integrated Fermi function. \textit{g} is the weak coupling constant which takes the value $g_V$ or $g_A$ according to whether the $\Delta J^{\pi}$ transition is associated with the vector or axial-vector weak interaction. The dynamics part of the rate equation
is given by

\begin{eqnarray}\label{e}
	B_{ij}(\Delta
	J^{\pi})=\frac{1}{12}\zeta^{2}(w_{m}^{2}-1)-\frac{1}{6}\zeta^{2}w_{m}w+\frac{1}{6}\zeta^{2}w^{2},
\end{eqnarray}
where $\zeta$ is
\begin{eqnarray}\label{efg}
	\zeta=2g_{A}\frac{\langle
		f|\vert\sum_{k}r_{k}[\textbf{C}^{k}_{1}\times
		{\sigma}]^{2}{\textbf{t}}^{k}_{-}\vert|i\rangle}{\sqrt{2J_{i}+1}},
\end{eqnarray}
and
\begin{eqnarray}\label{gkl}
	\textbf{C}_{lm}=\sqrt{{4\pi}{(2l+1)}^{-1}} \textbf{Y}_{lm},
\end{eqnarray}
the $\textbf{Y}_{lm}$ represents the spherical harmonics.

The kinematics portion of Eq.~(\ref{lij}) can be estimated using
\begin{eqnarray}\label{f}
	f_{ij} = \int_{1}^{w_{m}} w (w_{m}-w)^{2}({w^{2}-1})^{1/2}
	[(w_{m}-w)^{2}F_{1}(w, Z) \nonumber\\
	+ (w^{2}-1)F_{2}(w, Z)] (1-D_{-}) dw.
\end{eqnarray}
The term $w$ in the equation indicates the total energy of the electron (kinetic + rest mass energy). The total energy for $\beta$-decay is given by $ w_{m} =
E_{i}-E_{j} + m_{p} - m_{d}$, where $E_{i}$ ($E_{j}$) and $m_{p}$ ($m_{d}$) are excitation energy and mass of the parent (daughter) nucleus, respectively. $D_{-}$ are the electron distribution functions and are given by 
\begin{equation}\label{Gm}
	D_{-} = [exp (\frac{E-E_{f}}{kT})+1]^{-1},
\end{equation}
where $E_{f}$ and $E=(w-1)$ are the Fermi energy and kinetic energy of the electrons. The Fermi functions $F_{1}(\pm Z,w)$ and $F_{2}(\pm Z,w)$ appearing in Eq.~(\ref{f}) were calculated using the recipe of \cite{Gov71}.

Due to the prevailing high temperatures in the stellar core, there is a finite probability of occupation of parent excited states in the interior of massive stars. Hence $\beta$-decays
have a finite contribution from these parent excited states. Assuming thermal equilibrium, the probability of occupation of parent state $i$ may be estimated using 

\begin{equation}\label{pi}
	P_{i} = \frac {exp(-E_{i}/kT)}{\sum_{i=1}exp(-E_{i}/kT)}.
\end{equation}
The total stellar $\beta$-decay rates were finally calculated using
\begin{equation}\label{lb}
	\lambda^{\beta} = \sum_{ij}P_{i} \lambda_{ij}^{\beta}.
\end{equation}
The summation runs upon all parent and daughter states until satisfactory convergence in rate calculation
was obtained. A similar calculation was performed to compute continuum positron capture rates in high temperature-density environment.

In the pn-QRPA (N) model it was further
assumed that all
daughter excited states having energy larger than the
neutron separation energy ($S_{n}$), decayed through process of neutron emission.
The energy rate for neutron emission from daughter states was
computed using
\begin{equation}\label{ln}
	\lambda^{n} = \sum_{ij}P_{i}\lambda_{ij}(E_{j}-S_{n}),
\end{equation}
for all $E_{j} > S_{n}$. The probability of $\beta$-delayed neutron
emission, $P_{n}$, was calculated using
\begin{equation}\label{pn}
	P_{n} =
	\frac{\sum_{ij\prime}P_{i}\lambda_{ij\prime}}{\sum_{ij}P_{i}\lambda_{ij}},
\end{equation}
where $j\prime$ indicates the energy levels of the daughter nucleus
with $E_{j\prime}
> S_{n}$.
The $\lambda_{ij(\prime)}$ in Eq.~(\ref{ln}) and Eq.~(\ref{pn}),
represents the sum of positron capture and  $\beta$-decay rates,
for transition  from $i$ $\rightarrow$ $j(j\prime)$ state.
\subsubsection{Extension for odd-A Nuclei}

An extension of the pn-QRPA model is straight forward to
GT transitions from nuclear excited states. The excited states are constructed as phonon-correlated multi-quasiparticle states. The transition amplitudes between the multi-quasiparticle states were reduced to those of one-quasiparticle states. \\
For a nucleus with an odd nucleon, i.e., a proton and/or a neutron, low-lying states were obtained by lifting the quasiparticle in the orbit of the smallest energy to higher-lying orbits. States of an odd-proton even-neutron nucleus were expressed by three-proton states or one-proton two-neutron states, corresponding to excitation of a proton or a neutron,
\begin{eqnarray}\label{opn1}
		|p_1p_2p_{3corr}\rangle =a^\dagger_{p_1}a^\dagger_{p_2}a^\dagger_{p_3}|-\rangle \nonumber \\
		~~~~~~~~~~~~~~~+ \frac{1}{2}\sum_{p^{'}_1p^{'}_2n^{'}\omega}a^\dagger_{p^{'}_1}a^\dagger_{p^{'}_2}a^\dagger_{n^{'}}A^\dagger_{\omega}(\mu)|-\rangle \nonumber \\
		~~~~~~~~~~~~~~~ \times \langle-|[a^\dagger_{p^{'}_1}a^\dagger_{p^{'}_2}a^\dagger_{n^{'}}A^\dagger_{\omega}(\mu)]^{\dagger}H_{31}a^\dagger_{p_1}a^\dagger_{p_2}a^\dagger_{p_3}|-\rangle \nonumber \\
		~~~~~~~~~~~~~~~ \times E_{p_1p_2p_3}(p^{'}_1p^{'}_2n^{'},\omega) 
\end{eqnarray} 

\begin{eqnarray}\label{opn2}
	|p_1n_1n_{2corr}\rangle =a^\dagger_{p_1}a^\dagger_{n_1}a^\dagger_{n_2}|-\rangle \nonumber \\
	~~~~~~~~~~~~~~~+ \frac{1}{2}\sum_{p^{'}_1p^{'}_2n^{'}\omega}a^\dagger_{p^{'}_1}a^\dagger_{p^{'}_2}a^\dagger_{n^{'}}A^\dagger_{\omega}(-\mu)|-\rangle \nonumber \\
	~~~~~~~~~~~~~~~ \times \langle-|[a^\dagger_{p^{'}_1}a^\dagger_{p^{'}_2}a^\dagger_{n^{'}}A^\dagger_{\omega}(-\mu)]^{\dagger}H_{31}a^\dagger_{p_1}a^\dagger_{n_1}a^\dagger_{n_2}|-\rangle \nonumber \\
	~~~~~~~~~~~~~~~ \times E_{p_1n_1n_2}(p^{'}_1p^{'}_2n^{'},\omega) \nonumber \\
	~~~~~~~~~~~~~~~+\frac{1}{6}\sum_{n^{'}_1n^{'}_2n^{'}_3\omega}a^\dagger_{n^{'}_1}a^\dagger_{n^{'}_2}a^\dagger_{n^{'}_3}A^\dagger_{\omega}(\mu)|-\rangle \nonumber \\
	~~~~~~~~~~~~~~~ \times \langle-|[a^\dagger_{n^{'}_1}a^\dagger_{n^{'}_2}a^\dagger_{n^{'}_3}A^\dagger_{\omega}(\mu)]^{\dagger}H_{31}a^\dagger_{p_1}a^\dagger_{n_1}a^\dagger_{n_2}|-\rangle \nonumber \\
	~~~~~~~~~~~~~~~ \times E_{p_1n_1n_2}(n^{'}_1n^{'}_2n^{'}_3,\omega)
\end{eqnarray} 
with the energy denominators of first order perturbation,
\begin{eqnarray}
	E_{abc}(\textit{def}, \omega)=\frac{1}{(\epsilon_a + \epsilon_b + \epsilon_c)-(\epsilon_d + \epsilon_e + \epsilon_f + \omega)}
\end{eqnarray}
Three-quasiparticle states of an even-proton odd-neutron nucleus were obtained from
Eqs.~\ref{opn1} and ~\ref{opn2} by the exchange of proton states and neutron states, p $\leftrightarrow$ n,
and $A^\dagger_{\omega}(\mu)$ $\leftrightarrow$ $A^\dagger_{\omega}(-\mu)$.
Amplitudes of the quasiparticle transitions between the three-quasiparticle states were reduced to those for correlated one-quasiparticle states. For parent nuclei with an odd-proton and even-neutron,
\begin{eqnarray}
	\langle p^{'}_1p^{'}_2n^{'}_{1corr}|t_{\pm}\sigma_{-\mu}|p_1p_2p_{3corr}\rangle \nonumber\\
	~~=\delta(p^{'}_1,p_2)\delta(p^{'}_2,p_3)\langle n^{'}_{1corr}|t_{\pm}\sigma_{-\mu}|p_{1corr}\rangle \nonumber\\
	~~-\delta(p^{'}_1,p_1)\delta(p^{'}_2,p_3)\langle n^{'}_{1corr}|t_{\pm}\sigma_{-\mu}|p_{2corr}\rangle \nonumber\\
	~~+\delta(p^{'}_1,p_1)\delta(p^{'}_2,p_2)\langle n^{'}_{1corr}|t_{\pm}\sigma_{-\mu}|p_{3corr}\rangle,
\end{eqnarray}
\begin{eqnarray}
	\langle p^{'}_1p^{'}_2n^{'}_{1corr}|t_{\pm}\sigma_{\mu}|p_1n_1n_{2corr}\rangle \nonumber\\
	~~=\delta(n^{'}_1,n_2)[\delta(p^{'}_1,p_1)\langle p^{'}_{2corr}|t_{\pm}\sigma_{\mu}|n_{1corr}\rangle \nonumber\\
	~~-\delta(p^{'}_2,p_1)\langle p^{'}_{1corr}|t_{\pm}\sigma_{\mu}|n_{1corr}\rangle] \nonumber\\
	~~-\delta(n^{'}_1,n_1)[\delta(p^{'}_1,p_1)\langle p^{'}_{2corr}|t_{\pm}\sigma_{\mu}|n_{2corr}\rangle \nonumber\\
	~~-\delta(p^{'}_2,p_1)\langle p^{'}_{1corr}|t_{\pm}\sigma_{\mu}|n_{2corr}\rangle],
\end{eqnarray}
\begin{eqnarray}
	\langle n^{'}_1n^{'}_2n^{'}_{3corr}|t_{\pm}\sigma_{-\mu}|p_1n_1n_{2corr}\rangle \nonumber\\
	~~=\delta(n^{'}_2,n_1)\delta(n^{'}_3,n_2)\langle n^{'}_{1corr}|t_{\pm}\sigma_{-\mu}|p_{1corr}\rangle \nonumber\\
	~~-\delta(n^{'}_1,n_1)\delta(n^{'}_3,n_2)\langle n^{'}_{2corr}|t_{\pm}\sigma_{-\mu}|p_{1corr}\rangle \nonumber\\
	~~+\delta(n^{'}_1,n_1)\delta(n^{'}_2,n_2)\langle n^{'}_{3corr}|t_{\pm}\sigma_{-\mu}|p_{1corr}\rangle,
\end{eqnarray}
Similarly, for an odd-neutron even-proton nucleus,
\begin{eqnarray}
	\langle p^{'}_1n^{'}_1n^{'}_{2corr}|t_{\pm}\sigma_{\mu}|n_1n_2n_{3corr}\rangle \nonumber\\
	~~=\delta(n^{'}_1,n_2)\delta(n^{'}_2,n_3)\langle p^{'}_{1corr}|t_{\pm}\sigma_{\mu}|n_{1corr}\rangle \nonumber\\
	~~-\delta(n^{'}_1,n_1)\delta(n^{'}_2,n_3)\langle p^{'}_{1corr}|t_{\pm}\sigma_{\mu}|n_{2corr}\rangle \nonumber\\
	~~+\delta(n^{'}_1,n_1)\delta(n^{'}_2,n_2)\langle p^{'}_{1corr}|t_{\pm}\sigma_{\mu}|n_{3corr}\rangle,
\end{eqnarray}
\begin{eqnarray}
	\langle p^{'}_1n^{'}_1n^{'}_{2corr}|t_{\pm}\sigma_{-\mu}|p_1p_2n_{1corr}\rangle \nonumber\\
	~~=\delta(p^{'}_1,p_2)[\delta(n^{'}_1,n_1)\langle n^{'}_{2corr}|t_{\pm}\sigma_{-\mu}|p_{1corr}\rangle \nonumber\\
	~~-\delta(n^{'}_2,n_1)\langle n^{'}_{1corr}|t_{\pm}\sigma_{-\mu}|p_{1corr}\rangle] \nonumber\\
	~~-\delta(p^{'}_1,p_1)[\delta(n^{'}_1,n_1)\langle n^{'}_{2corr}|t_{\pm}\sigma_{-\mu}|p_{2corr}\rangle \nonumber\\
	~~-\delta(n^{'}_2,n_1)\langle n^{'}_{1corr}|t_{\pm}\sigma_{-\mu}|p_{2corr}\rangle],
\end{eqnarray}
\begin{eqnarray}
	\langle p^{'}_1p^{'}_2p^{'}_{3corr}|t_{\pm}\sigma_{\mu}|p_1p_2n_{1corr}\rangle \nonumber\\
	~~=\delta(p^{'}_2,p_1)\delta(p^{'}_3,p_2)\langle p^{'}_{1corr}|t_{\pm}\sigma_{\mu}|n_{1corr}\rangle \nonumber\\
	~~-\delta(p^{'}_1,p_1)\delta(p^{'}_3,p_2)\langle p^{'}_{2corr}|t_{\pm}\sigma_{\mu}|n_{1corr}\rangle \nonumber\\
	~~+\delta(p^{'}_1,p_1)\delta(p^{'}_2,p_2)\langle p^{'}_{3corr}|t_{\pm}\sigma_{\mu}|n_{1corr}\rangle,
\end{eqnarray}
For further details we refer to Ref. ~\cite{Mut92}. 
\section{Results and Discussion}

The terrestrial
$\beta$-decay half-lives, stellar weak rates, $\beta$-delayed neutron emission probabilities, phase space integrals and charge-changing strength distributions computed using the pn-QRPA (N) model include both allowed GT and U1F transitions. The pn-QRPA (N) model is well-known for its very good predictive power of estimating half-lives of unknown nuclei, especially for nuclei having smaller half-life values (nuclei far off from the line of stability)~\cite{Hir93, Nab16}.  
It is worth noting that no quenching factor was utilized in the present calculation. In the recent past, we have reported $\beta$-decay half-lives, GT strength distributions, phase space and stellar weak rates of 13 closed-shell waiting point nuclei ($N$ = 50, 82) employing the deformed pn-QRPA (N) model \cite{Nab19}. 



In this present project, we extend our calculation to $N$ = 126 waiting point nuclei and select 17 cases given in Table~\ref{tab1}. The calculated $\beta$-decay half-lives using the pn-QRPA (N) and  pn-QRPA (WS) models are shown in Table~\ref{tab1}. For the sake of comparison, previous calculations including allowed GT calculation of \cite{Mol97}, QRPA-FRDM \cite{Mol03}, CQRPA-DF3 \cite{Bor06,Bor11}, SM \cite{Suz12}, LSSM \cite{Zhi13}, empirical formula based calculation \cite{Zhou17} and NUBASE2016 values~\cite{Aud17} are also shown in the table. The calculation done by \cite{Mol97} includes only allowed GT contribution. QRPA-FRDM calculation comprises of both allowed GT and first-forbidden (FF) contributions where allowed GT part was computed via QRPA approximation and gross theory was used for the computation of FF part. It is to be noted that the experimental evaluations in Ref~\cite{Mol03} were taken from Refs.~\cite{Mol03Ev1,Pfe02}. The SM computed the half-lives (including allowed GT and FF transitions) employing a quenching factor of 0.7. The LSSM results contain allowed GT and FF rates including rank 0, 1 and 2 operator values. LSSM reported that FF rates provide a substantial reduction in the total calculated half-lives for the cases having N = $126$. It was reported in Ref.~\cite{Bor06} that FF contribution dominates over allowed GT  for $Z \geq 76$.  It is to be noted that LSSM incorporated varying quenching values ranging from 0.38 to 1.266, whereas the pn-QRPA (N) and pn-QRPA (WS) models did not incorporate any explicit quenching factor as mentioned earlier. The LSSM calculated half-life values are smaller than pn-QRPA values. A fully converged computation of the FF transition strength in LSSM approach was denied due to computational constraints. They applied the Lanczos method to derive the strength but was again limited up to 100 iterations which were insufficient for converging the states above 2.5 \textit{MeV} excitation energies. The pn-QRPA approach had no such limitations and was able to converge the states for excitation energies well over 10 \textit{MeV}. 
It may be noted from Table~\ref{tab1} that the pn-QRPA (N) and pn-QRPA (WS) calculated half-lives are in decent comparison with the measured data. The FF (U1F) component significantly reduced the computed half-lives, efficiently for higher Z numbers, and improved the comparison with  experimental data (at times also with the previous calculations). 


Table~\ref{tab2} and Table~\ref{tab3} display the calculated $\beta$-decay (electron emission) rates (allowed GT and U1F) for the selected $N$ = 126 isotones. The rates are shown at stellar temperatures ($T_{9}$ = 1, 2, 5, 15 \& 25) $\textit{GK}$ and densities ($\rho$$\it Y_{e}$=10$^{2}$), ($\rho$$\it Y_{e}$=10$^{6}$) and ($\rho$$\it Y_{e}$=10$^{10}$) \textit{gcm$^{-3}$}. The rates are tabulated in log to base 10 values (in units of \textit{s$^{-1}$}). These rates were calculated for a broad range of temperature ($T_{9}$ = 0.01--30) \textit{GK} including the typical range of temperature for $r$-process ($T_{9}$ = 1--3) \textit{GK}. The tables show that the $\beta$-decay rates on the selected nuclei increase with a rise in the stellar temperature. This may be attributed to the fact that the occupation probability of parent excited states increases with temperature rise and thus contributes effectively to the total rates. It is found that the electron Fermi energy increases with an increase in the density of  stellar core which leads to a reduction in the available phase space and a corresponding decrease in the calculated $\beta$-decay rates.  





Table~\ref{tab6} shows estimated values of the $\beta$-delayed neutron emission probability for selected $N$ = 126 nuclei using different models. Previous computations including  QRPA-FRDM \cite{Mol03}, CQRPA-DF3 \cite{Bor06} and LSSM \cite{Zhi13} are also shown in Table~\ref{tab6}. The calculations are not in good agreement with each other.  The main reason for the varying predictions would be the calculation of daughter energy levels and computation of charge-changing matrix elements between the parent and daughter states in the different models. All the calculated probabilities reduce, generally, towards higher Z numbers.

\begin{figure*}
	\centering
	\includegraphics[width=1.\textwidth]{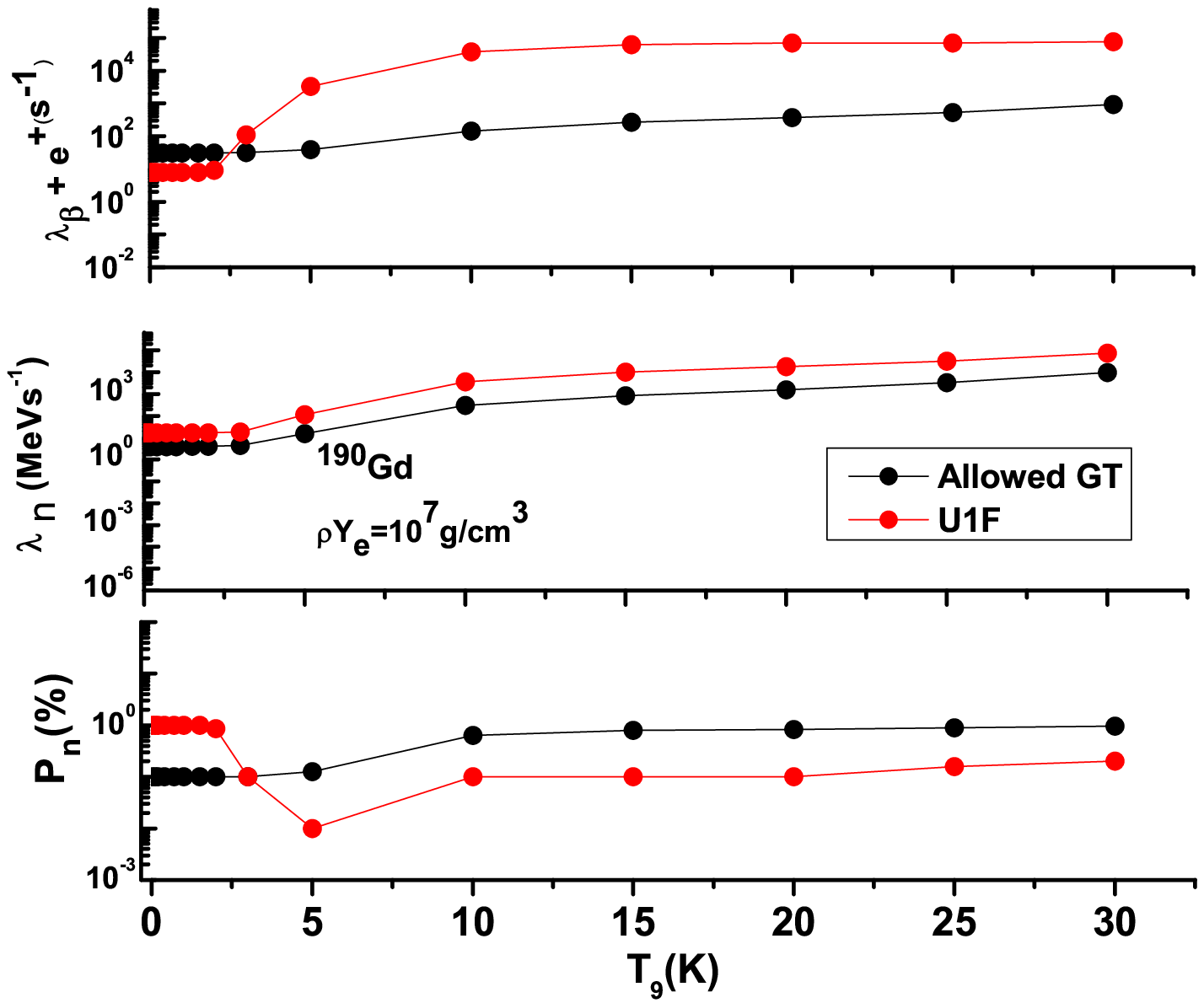}
	\caption{The pn-QRPA (N) computed $\beta$-decay and positron
		capture rates (upper panel), energy rates of $\beta$-delayed neutron
		(middle panel) and their emission probabilities (bottom panel) for
		$^{190}$Gd as a function of core temperature at fixed stellar density of
		10$^{7}$\textit{g.cm$^{-3}$}. The allowed GT and U1F contributions are shown
		separately.} \label{fig2}
\end{figure*}

\begin{figure*}
	\centering
	\includegraphics[width=1.\textwidth]{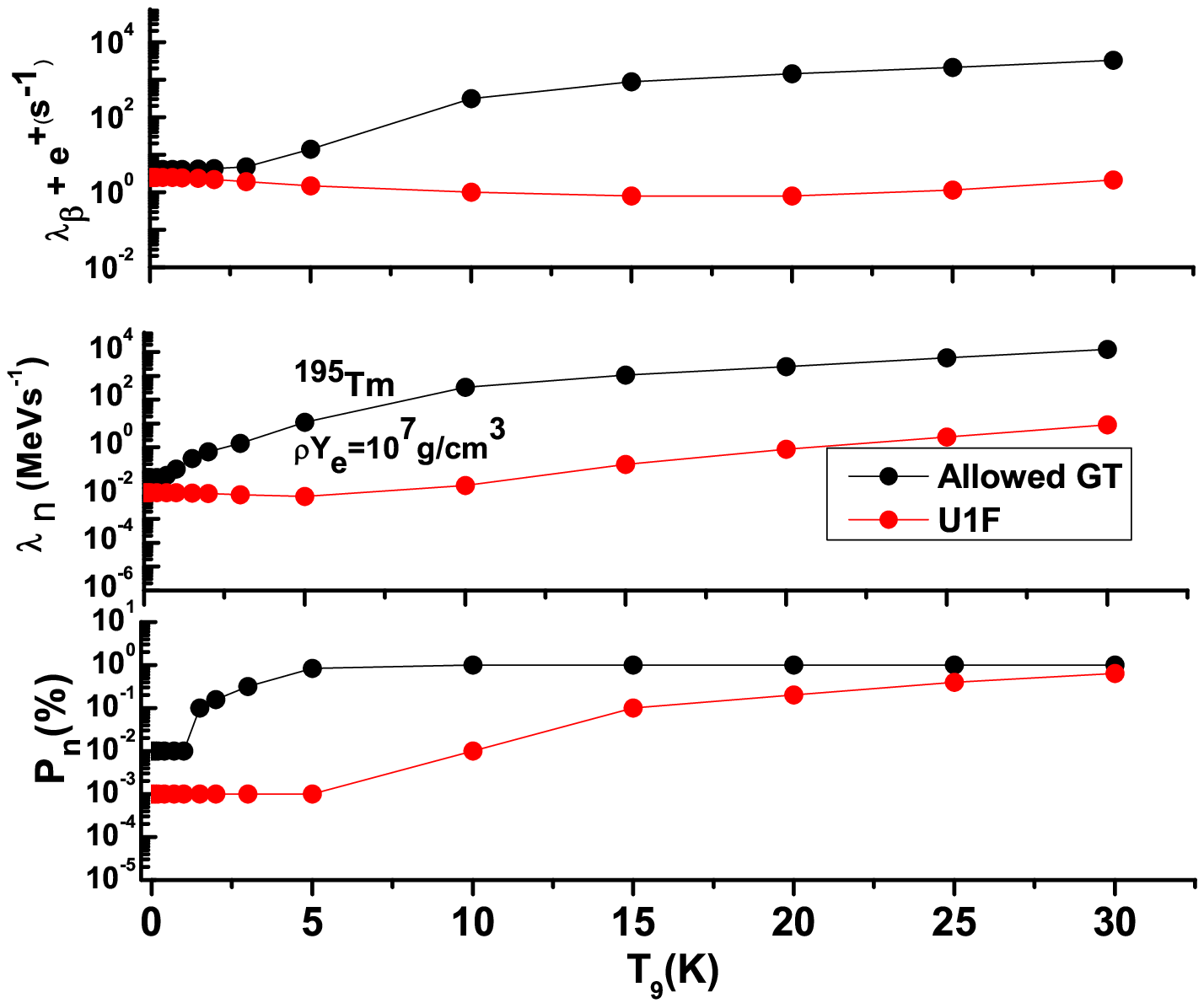}
	\caption{\scriptsize Same as Fig.~\ref{fig2} but for $^{195}$Tm.}
	\label{fig3}
\end{figure*}

\begin{figure*}
	\centering
	\includegraphics[width=1.\textwidth]{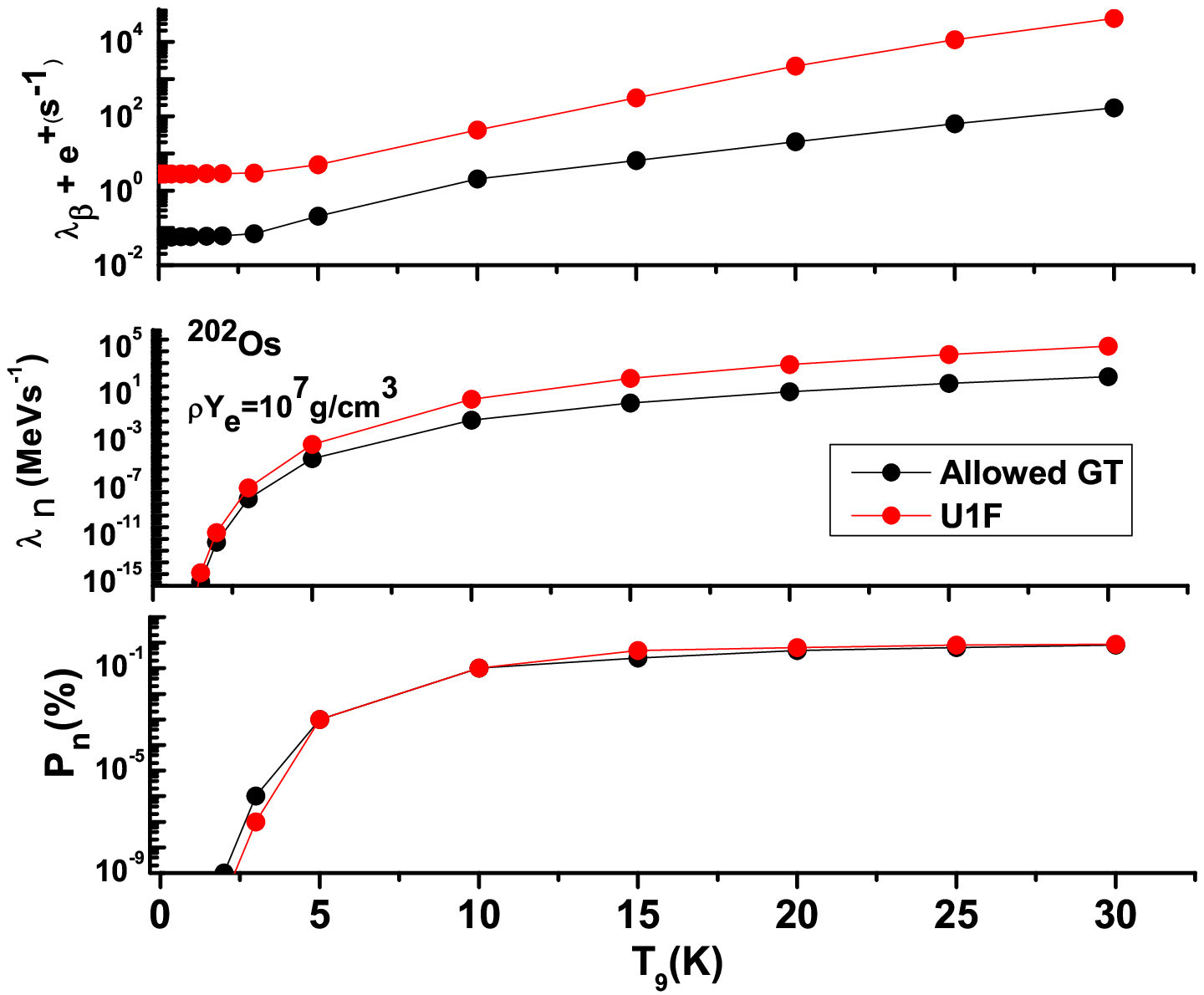}
	\caption{\scriptsize Same as Fig.~\ref{fig2} but for $^{202}$Os.}
	\label{fig4}
\end{figure*}

\begin{figure*}
	\centering
	\includegraphics[width=1.\textwidth]{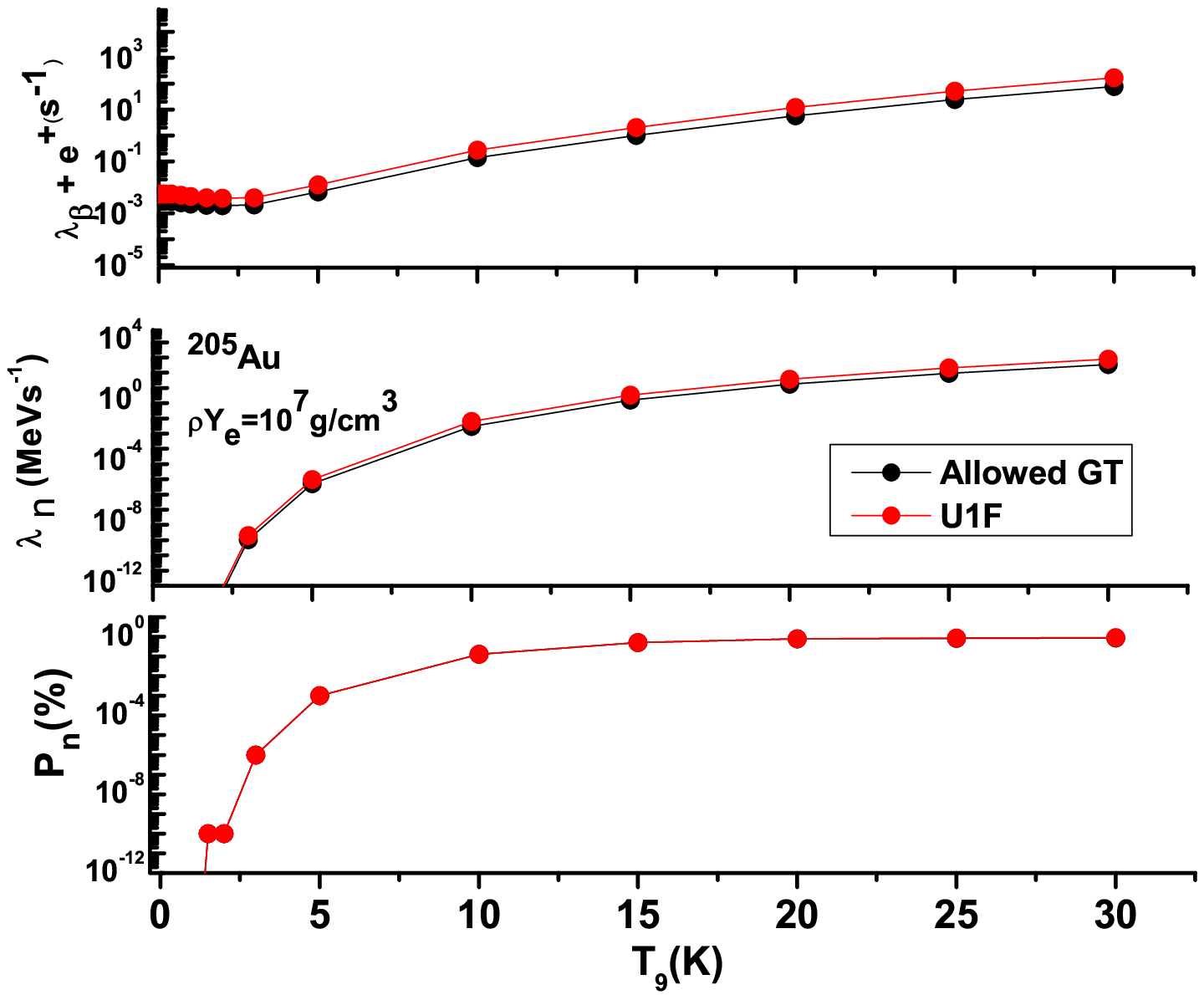}
	\caption{\scriptsize Same as Fig.~\ref{fig2} but for $^{205}$Au.}
	\label{fig5}
\end{figure*}

To investigate weak rates in stellar environment, we computed the $\beta$-decay  and (continuum) positron capture rates for a wide range of density (10 - 10$^{11}$) \textit{g/cm$^{3}$} and temperature ($T_{9}$ = 0.01--30) \textit{GK} for the selected waiting point nuclei.
Due to space considerations, we chose to display results of only four cases (two even-even and two odd-A nuclei). Figs.~(\ref{fig2}-~\ref{fig5}) display the computed weak rates for $^{190}$Gd, $^{195}$Tm,$^{202}$Os and $^{205}$Au, respectively. Each figure comprises of three panels. The upper panel displays the sum of $\beta$-decay and positron capture rates (in units of $s^{-1}$) as a function of stellar temperature. The middle panel depicts the computed $\beta$-delayed neutron energy rates (in units of $MeV.s^{-1}$) while the lower panel shows the calculated $\beta$-delayed neutron emission probability ($P$$_{n}$) values. A fixed core density of 10$^{7}$ \textit{g/cm$^{3}$}  was assumed for all these calculations. The allowed GT and U1F rates are shown separately in all these figures.


The allowed GT rates for $^{190}$Gd, shown in Fig.~\ref{fig2} (upper panel), are roughly factor 4 greater than the U1F rates at low stellar temperatures ($T_{9}$ = 0.01--2) \textit{GK}. However, the U1F rates increase more rapidly beyond ($T_{9}$ = 2) \textit{GK} and exceed the allowed GT rates, by up to a factor 66, at higher $T_{9}$ values.
The emission rate of $\beta$-delayed neutrons was found higher due to U1F transitions when compared against allowed GT transitions at  low-temperature range ($T_{9}$ = 0.01--3) \textit{GK}. Consequently, the energy rates (middle panel) due to U1F transitions are factor 4 higher than those due to allowed GT transitions at low $T_9$ values and roughly an order of magnitude bigger for higher temperatures. The corresponding  $\beta$-delayed neutron emission probabilities are roughly an order of magnitude greater due to U1F transitions alone at low stellar temperatures. The probability values due to allowed GT transitions exceed those due to U1F transitions at high $T_{9}$ values. This change may be ascribed to the behavior of the available phase spaces for allowed GT and U1F transitions, which we discuss later.  

In case of $^{195}$Tm, shown in Fig.~\ref{fig3}, the allowed GT rates are comparable to U1F rates at low temperatures and are orders of magnitude bigger than U1F rates for higher $T_{9}$ values ($T_{9}$ = 10--30) \textit{GK}. Similarly, the energy rates due to allowed GT transitions are up to an order bigger than those due to U1F at low temperatures ($T_{9}$ = 0.01--1) \textit{GK} while they are orders of magnitude bigger beyond ($T_{9}$ = 1) \textit{GK}. Correspondingly, the neutron emission probability values (bottom panel) due to allowed GT transitions were found up to an order (two orders) of magnitude bigger than U1F for low (high) $T_{9}$ values.

The $\beta$-decay rates and the energy rates of $\beta$-delayed neutron for the case of $^{202}$Os, shown in Fig.~\ref{fig4}, due to U1F transitions are greater than the respective rates due to allowed GT transitions by up to 2 orders of magnitude at all temperatures. The corresponding $\beta$-delayed neutron emission probabilities due to U1F transitions are smaller than those due to the only allowed GT at low-temperatures. The rates are comparable for higher $T_{9}$ values. 

Similarly, the $\beta$-decay and energy rates of $\beta$-delayed neutron for $^{205}$Au (Fig.~\ref{fig5}), due to U1F transitions, are up to a factor 2 bigger than the respective rates due to allowed GT transitions for all $T_{9}$ values. The probability values of $\beta$-delayed neutron emissions, both due to  allowed GT and U1F transitions are comparable for all temperature values.

The positron capture rates may be ignored as compared to $\beta$-decay rates for low temperature. They appear through $e^{-}$-- $e^{+}$
pair creation at temperature beyond  1 \textit{MeV} and compete well with $\beta$-decay rates at ($T_{9}$ = 30) \textit{GK}. In fact for some cases ($^{194}$Er, $^{196}$Yb, $^{202}$Os, $^{203}$Ir, $^{204}$Pt, $^{205}$Au and $^{206}$Hg) they are
bigger up to a factor 2 than the competing $\beta$-decay rates. The weak rates are the product of phase space and reduced transition probabilities. The behavior of weak rates shown in Figs.~(\ref{fig2}-~\ref{fig5}) may be traced back to
the strength distributions and phase space calculations which we discuss next.

\begin{figure*}
	\begin{tabular}{cc}
		\resizebox{0.5\hsize}{!}{\includegraphics*{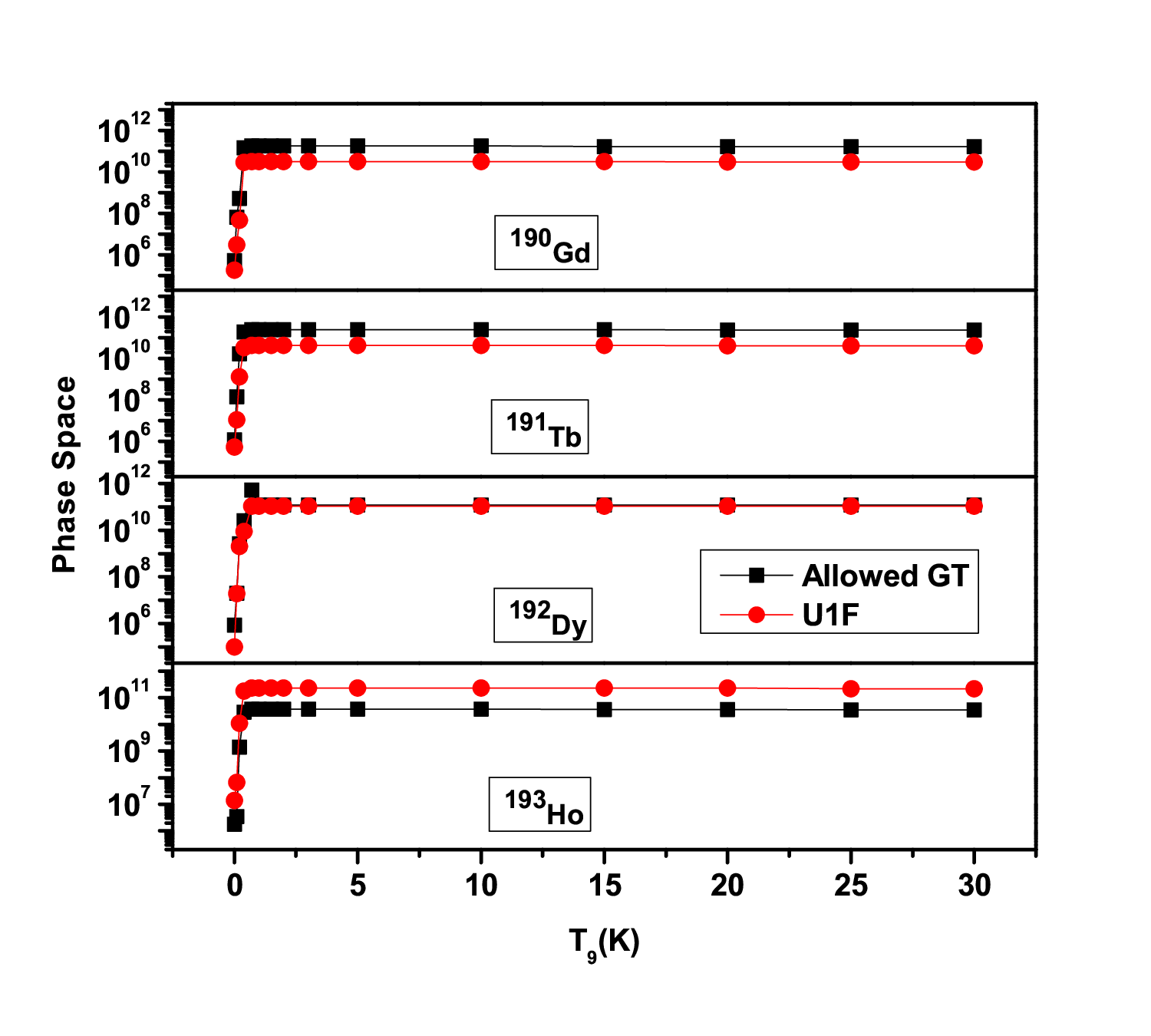}}&
		\resizebox{0.5\hsize}{!}{\includegraphics*{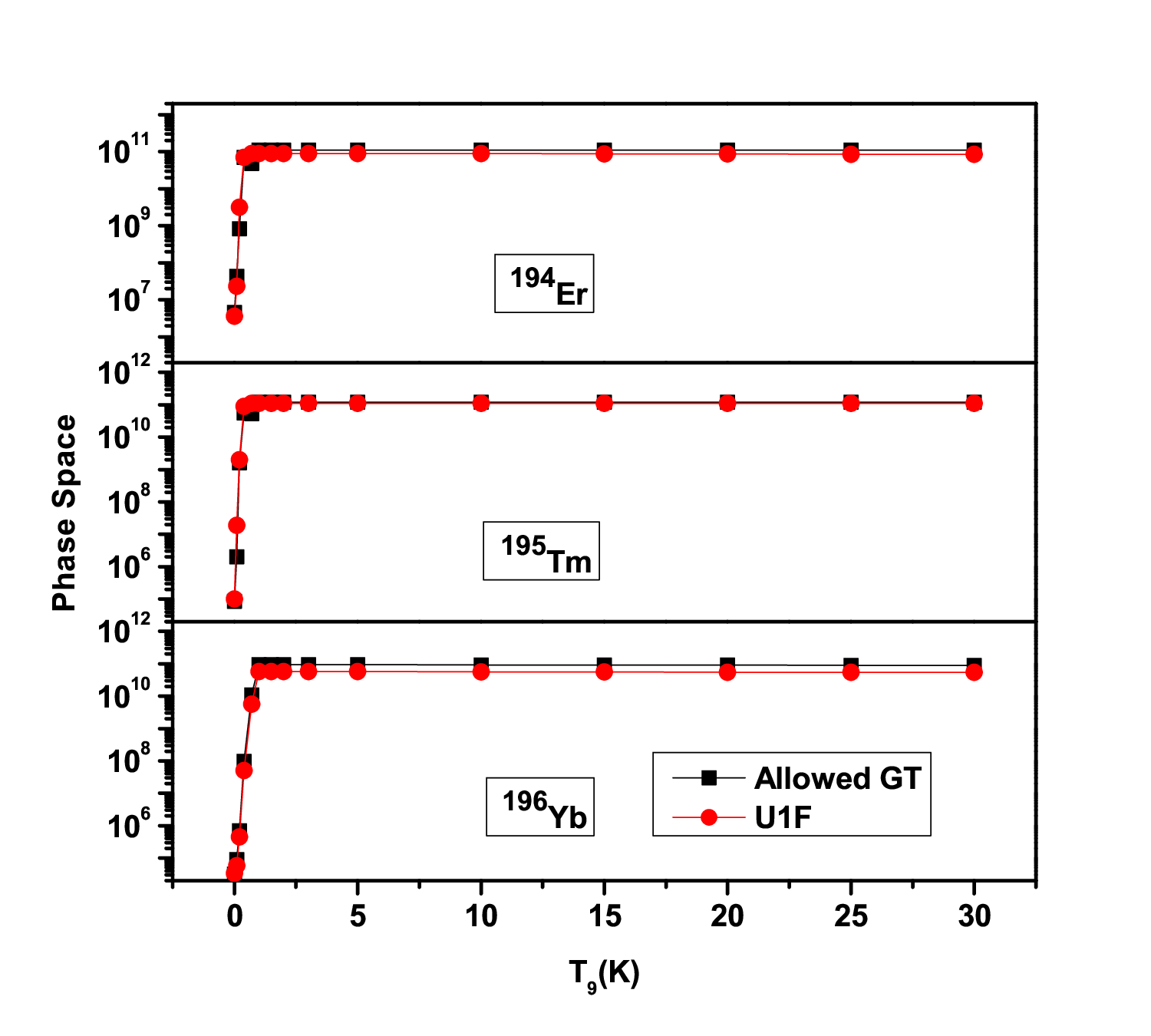}}\\
	\end{tabular}
	
	\caption{\scriptsize Computed phase space for $\beta$-decay (allowed GT and U1F) for $^{190}$Gd, $^{191}$Tb, $^{192}$Dy, $^{193}$Ho,
		$^{194}$Er, $^{195}$Tm and $^{196}$Yb as a function of stellar temperature at a 
		fixed density of 10$^{7}$\textit{g.cm$^{-3}$}.}\label{fig6} \centering
\end{figure*}
\begin{figure*}[h!t]
	\begin{tabular}{cc}
		\resizebox{0.5\hsize}{!}{\includegraphics*{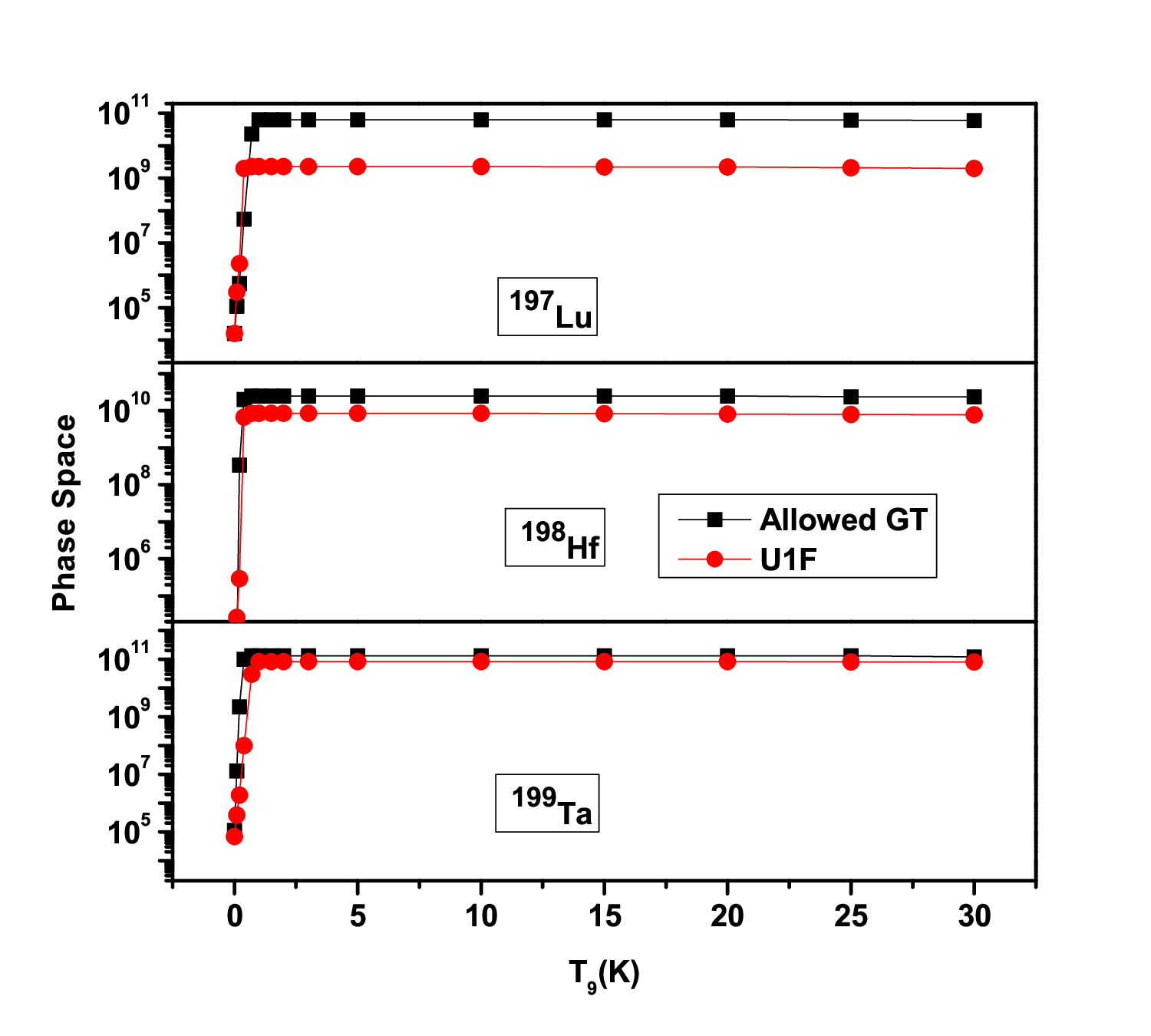}}&
		\resizebox{0.5\hsize}{!}{\includegraphics*{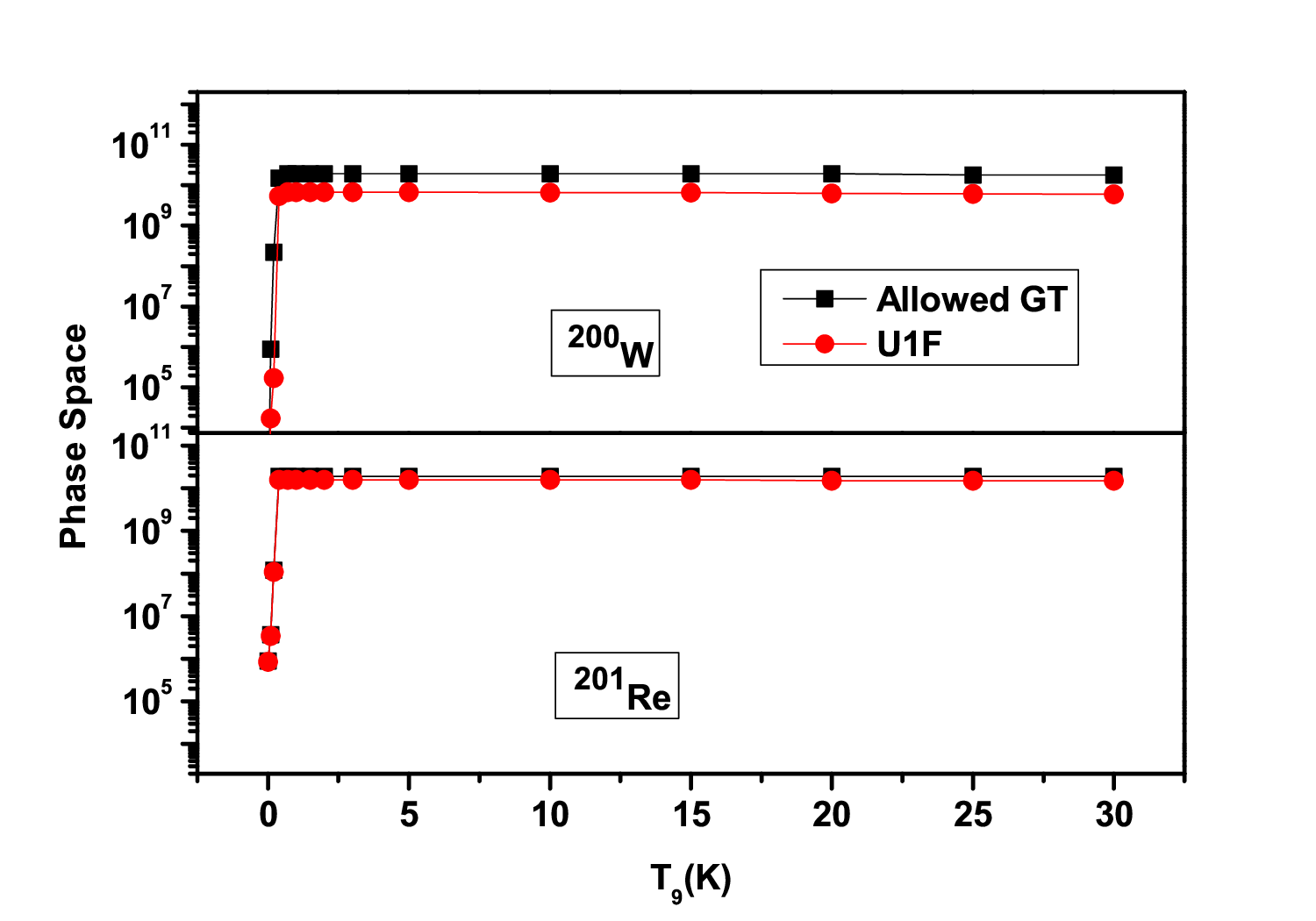}}\\
	\end{tabular}
	
	\caption{\scriptsize Same as Fig.~\ref{fig6} but for $^{197}$Lu, $^{198}$Hf, $^{199}$Ta,
		$^{200}$W and $^{201}$Re.}\label{fig7a}
	\centering
\end{figure*}
\begin{figure*}[h!t]
	\begin{tabular}{cc}
		\resizebox{0.5\hsize}{!}{\includegraphics*{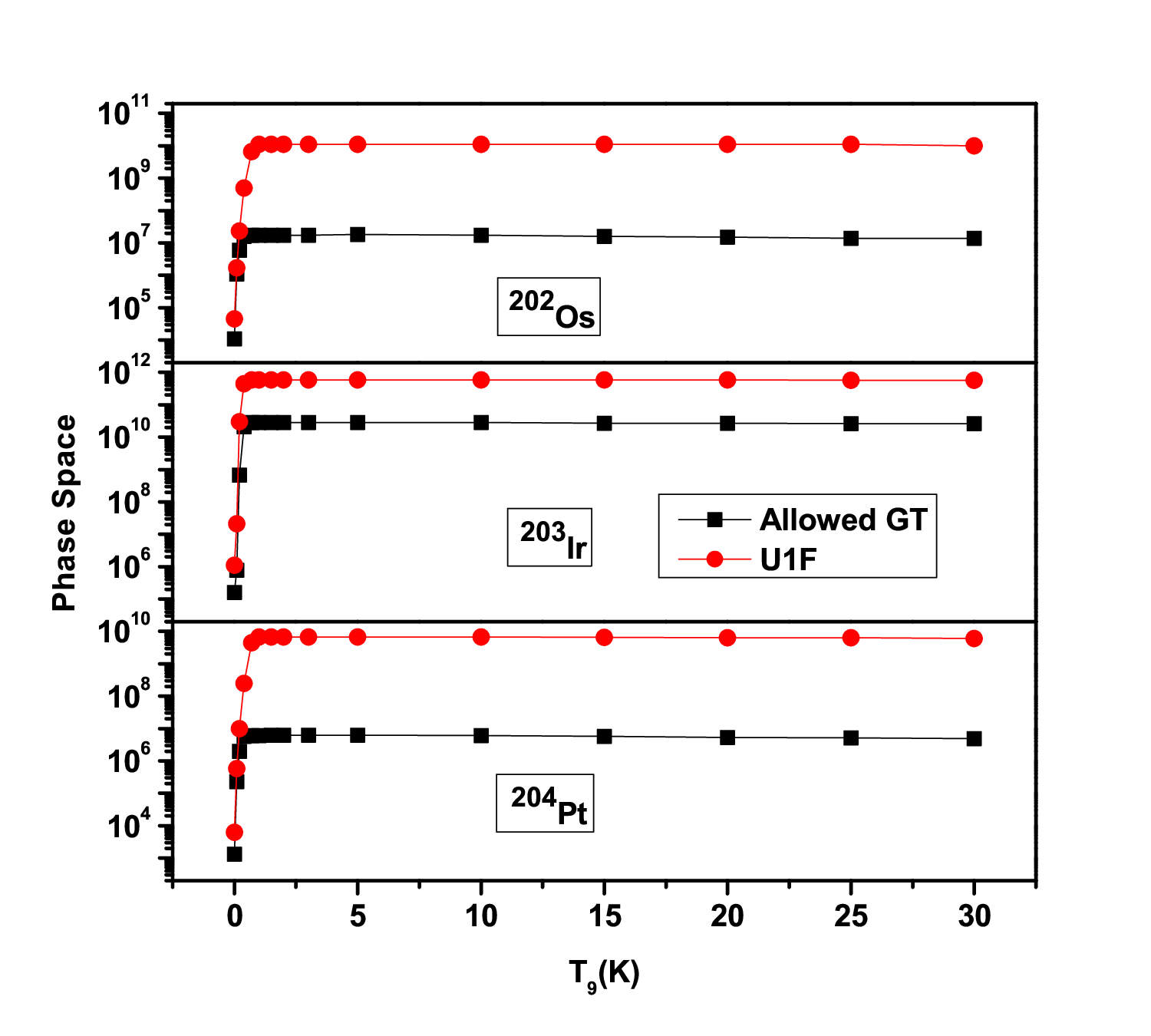}}&
		\resizebox{0.5\hsize}{!}{\includegraphics*{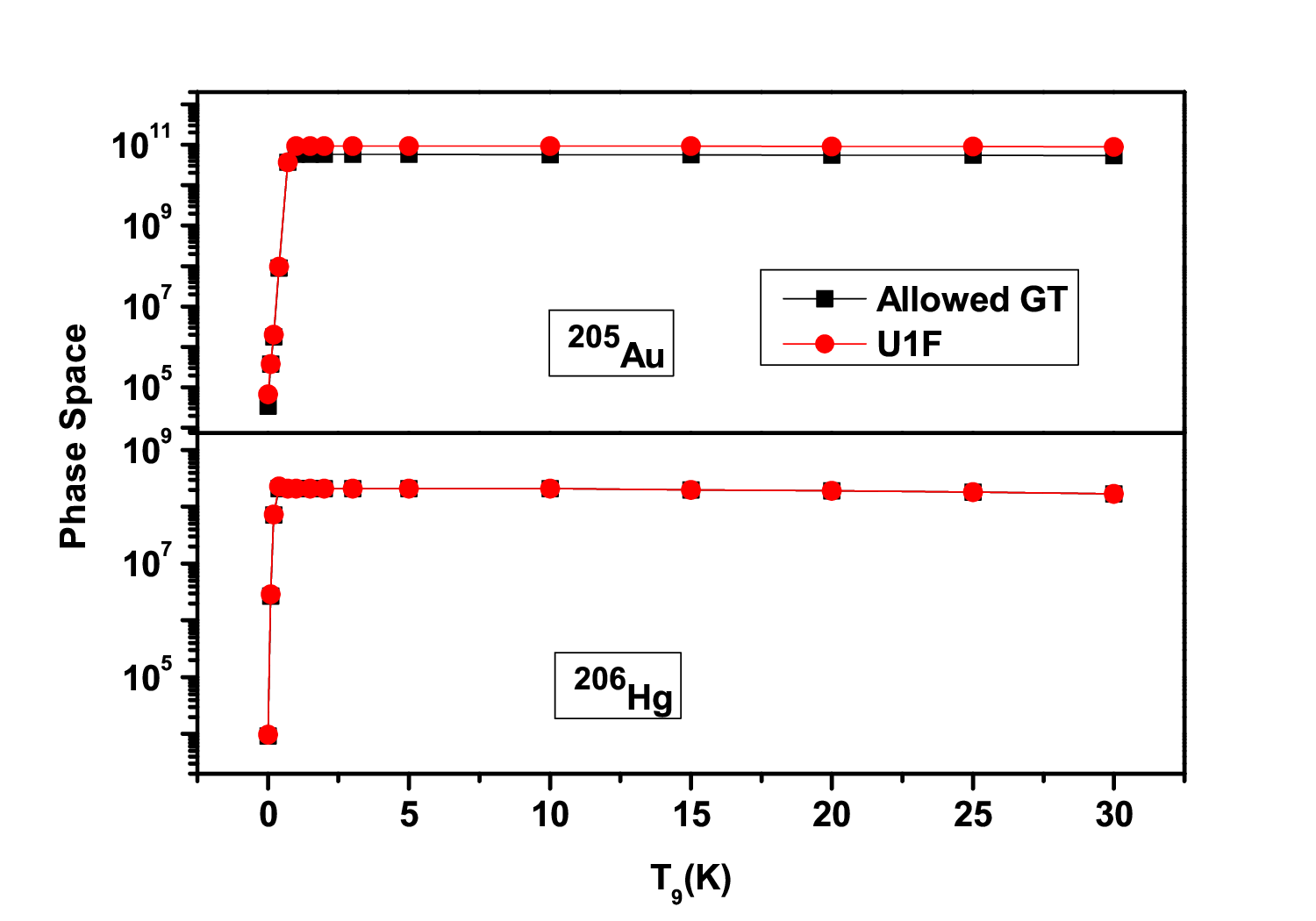}}\\
	\end{tabular}
	
	\caption{\scriptsize Same as Fig.~\ref{fig6} but for $^{202}$Os, $^{203}$Ir, $^{204}$Pt,
		$^{205}$Au and $^{206}$Hg.}\label{fig7}
	\centering
\end{figure*}

The computed phase space factors (allowed GT and U1F) at fixed density of 10$^{7}$ \textit{g/cm$^{3}$} as a function of stellar temperature, for the waiting point nuclei, are displayed in Figs.~(\ref{fig6}-\ref{fig7}). From these figures, it is noted that the allowed GT phase space is bigger than U1F phase space by up to an order of magnitude for the cases $^{190}$Gd, $^{191}$Tb, $^{192}$Dy, $^{194}$Er, $^{195}$Tm, $^{196}$Yb, $^{197}$Lu, $^{198}$Hf, $^{199}$Ta and $^{200}$W. The U1F phase is orders fo magnitude bigger than allowed GT phase space for $^{193}$Ho, $^{202}$Os, $^{203}$Ir, $^{204}$Pt and $^{205}$Au where as both the phase spaces are comparable for $^{201}$Re and $^{206}$Hg. Thus, we conclude that for all the cases shown in Fig.~\ref{fig6} and Fig.~\ref{fig7a}, the GT phase space is bigger than U1F (except for $^{193}$Ho) and the allowed GT contribution in reducing the half-lives is bigger than U1F contribution. The stellar rates due to GT transitions are also bigger for these cases (see Table \ref{tab2}).  For the cases shown in Fig.~\ref{fig7}, the U1F phase space is bigger than the GT phase space and contribute significantly to reducing half-lives and associated stellar rates. 

\begin{figure*}
	\begin{tabular}{cc}
		\resizebox{0.5\hsize}{!}{\includegraphics*{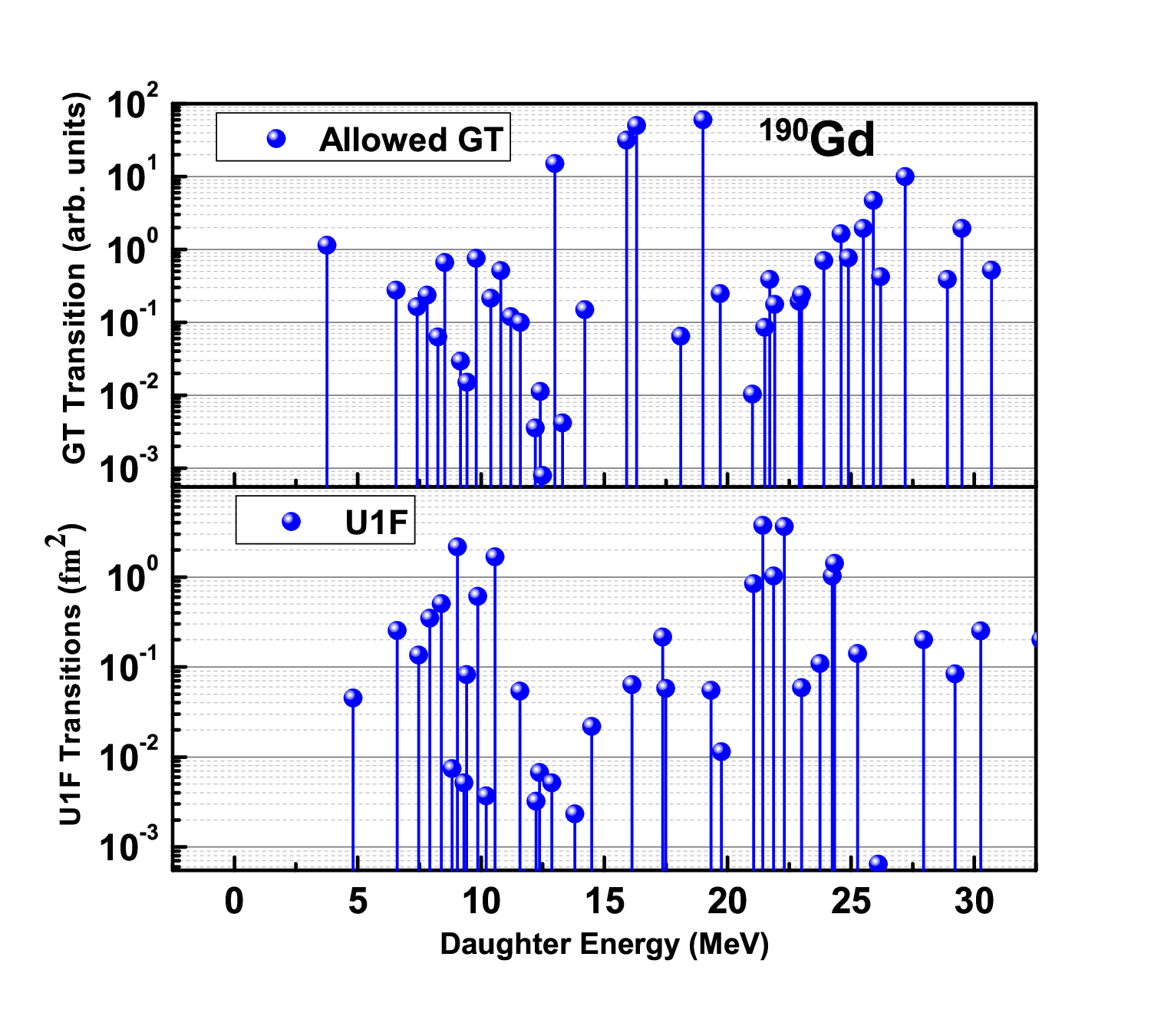}}&
		\resizebox{0.5\hsize}{!}{\includegraphics*{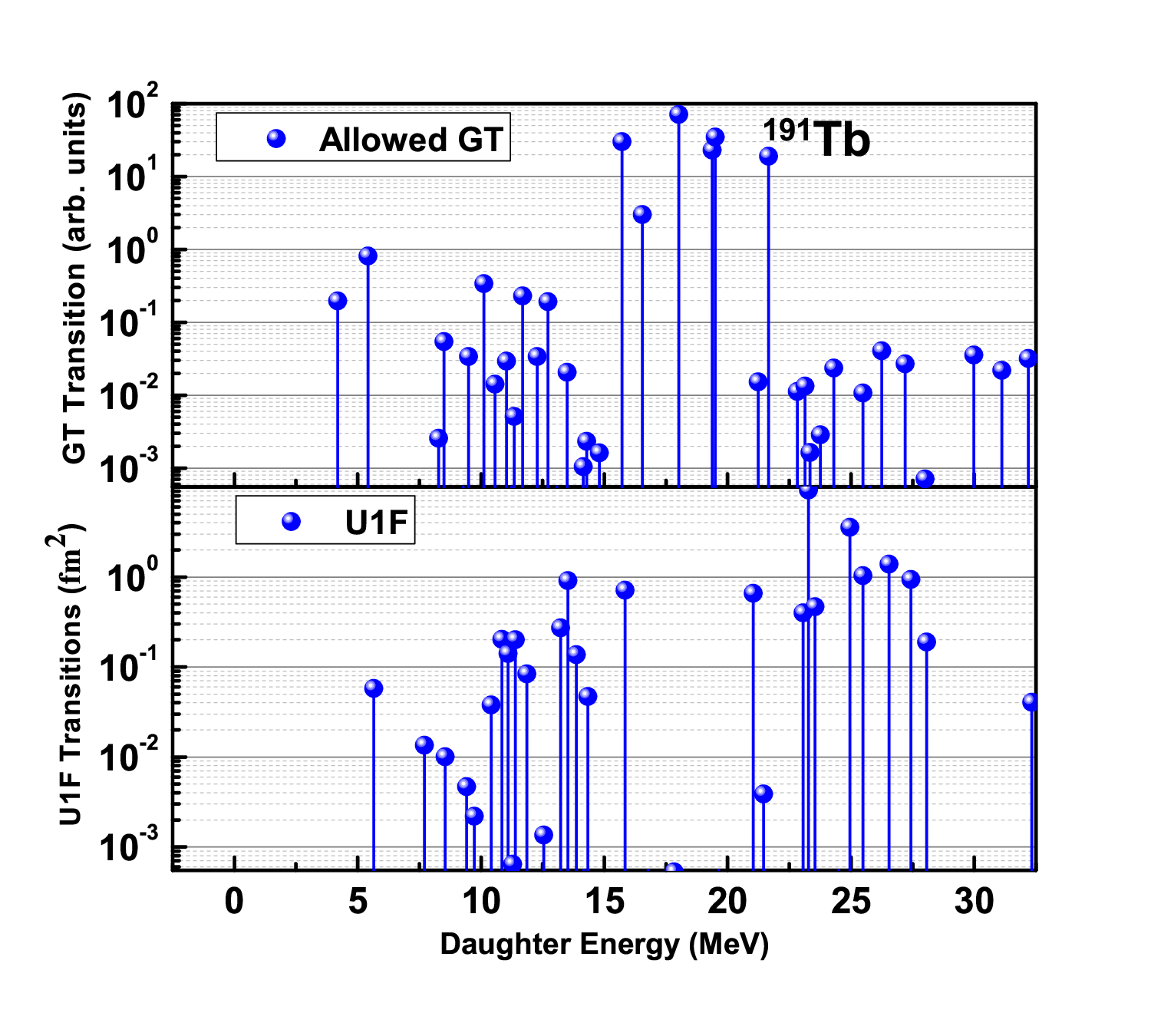}}\\
		\resizebox{0.5\hsize}{!}{\includegraphics*{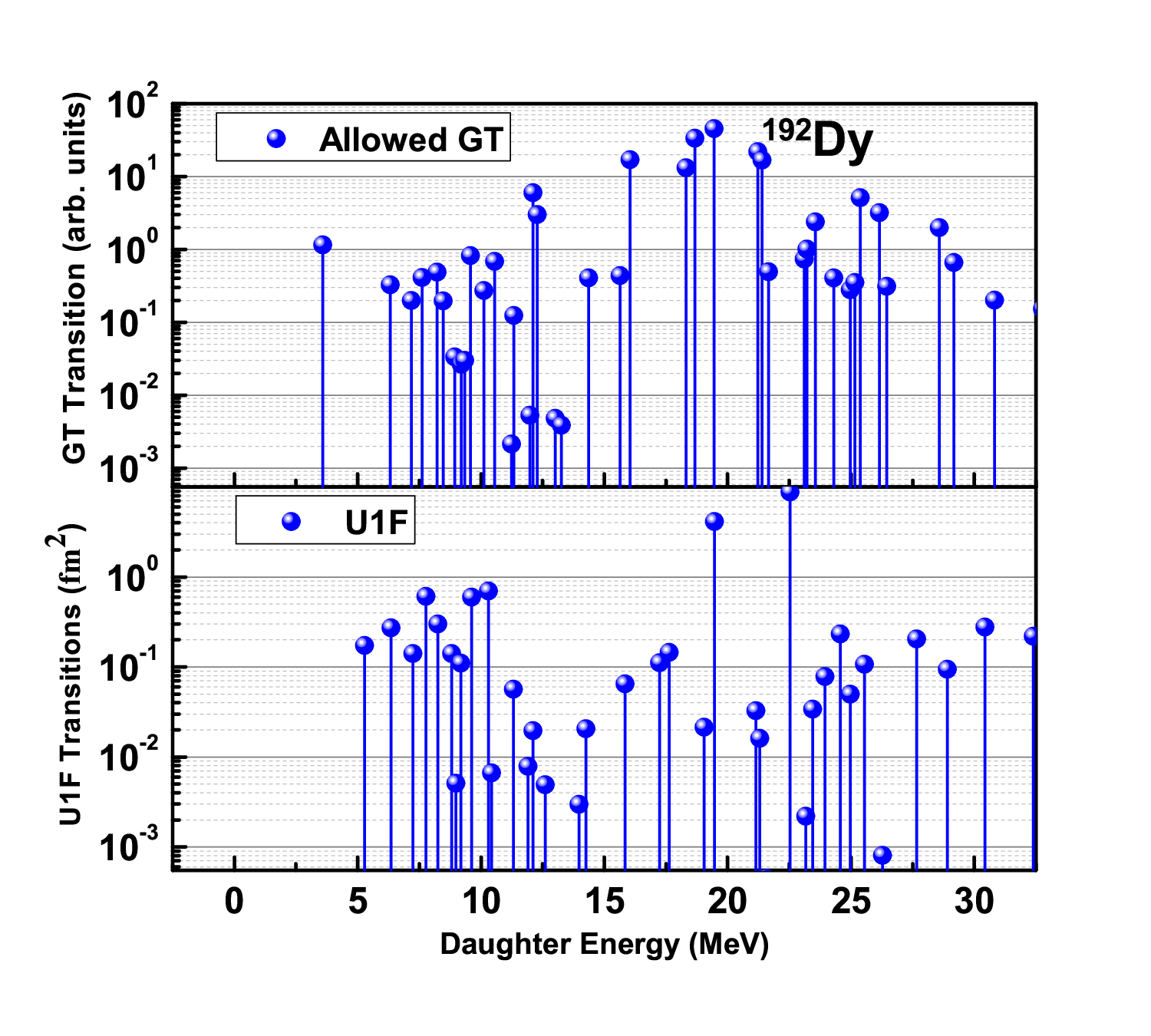}}&
		\resizebox{0.5\hsize}{!}{\includegraphics*{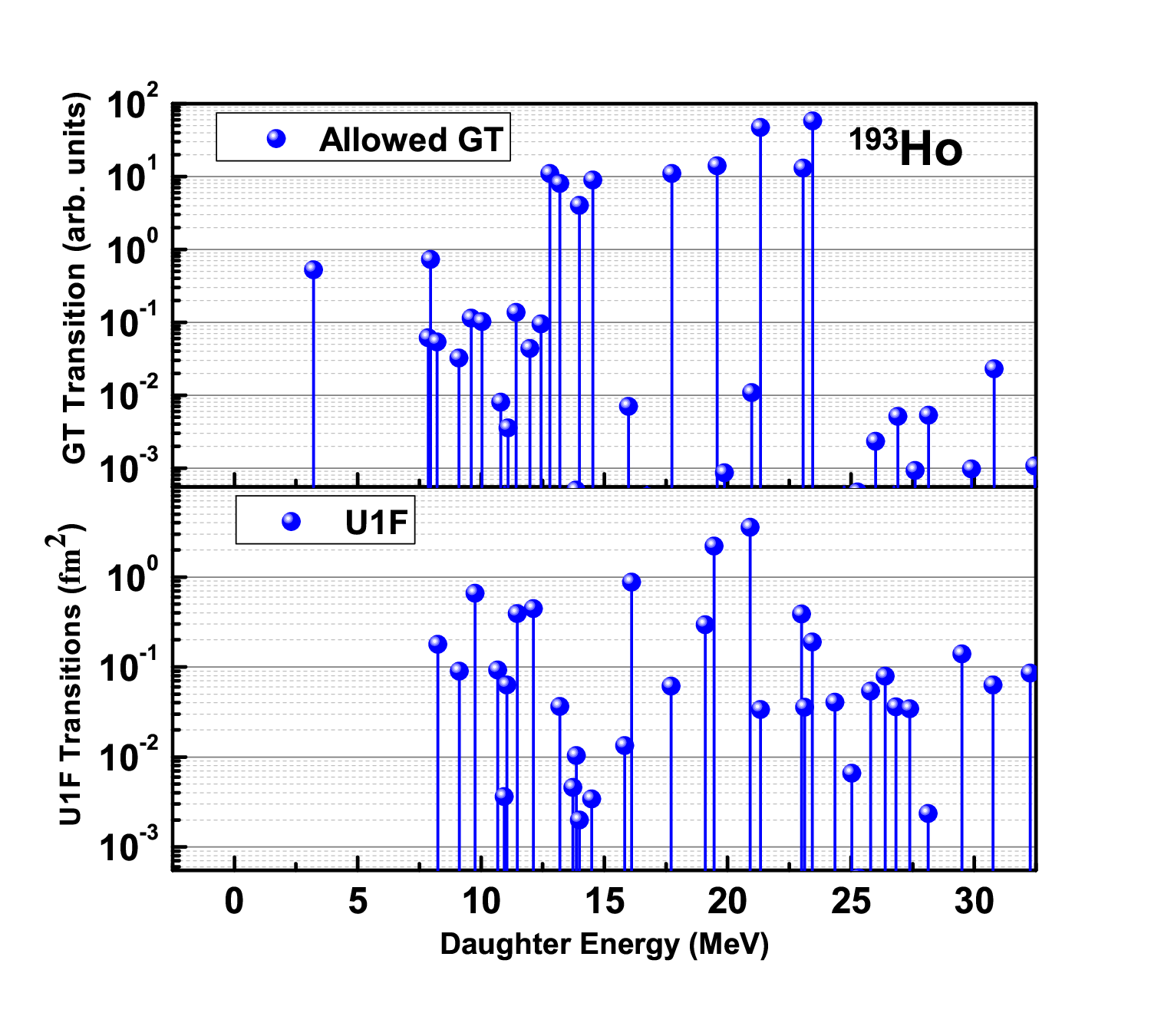}}\\
		\resizebox{0.5\hsize}{!}{\includegraphics*{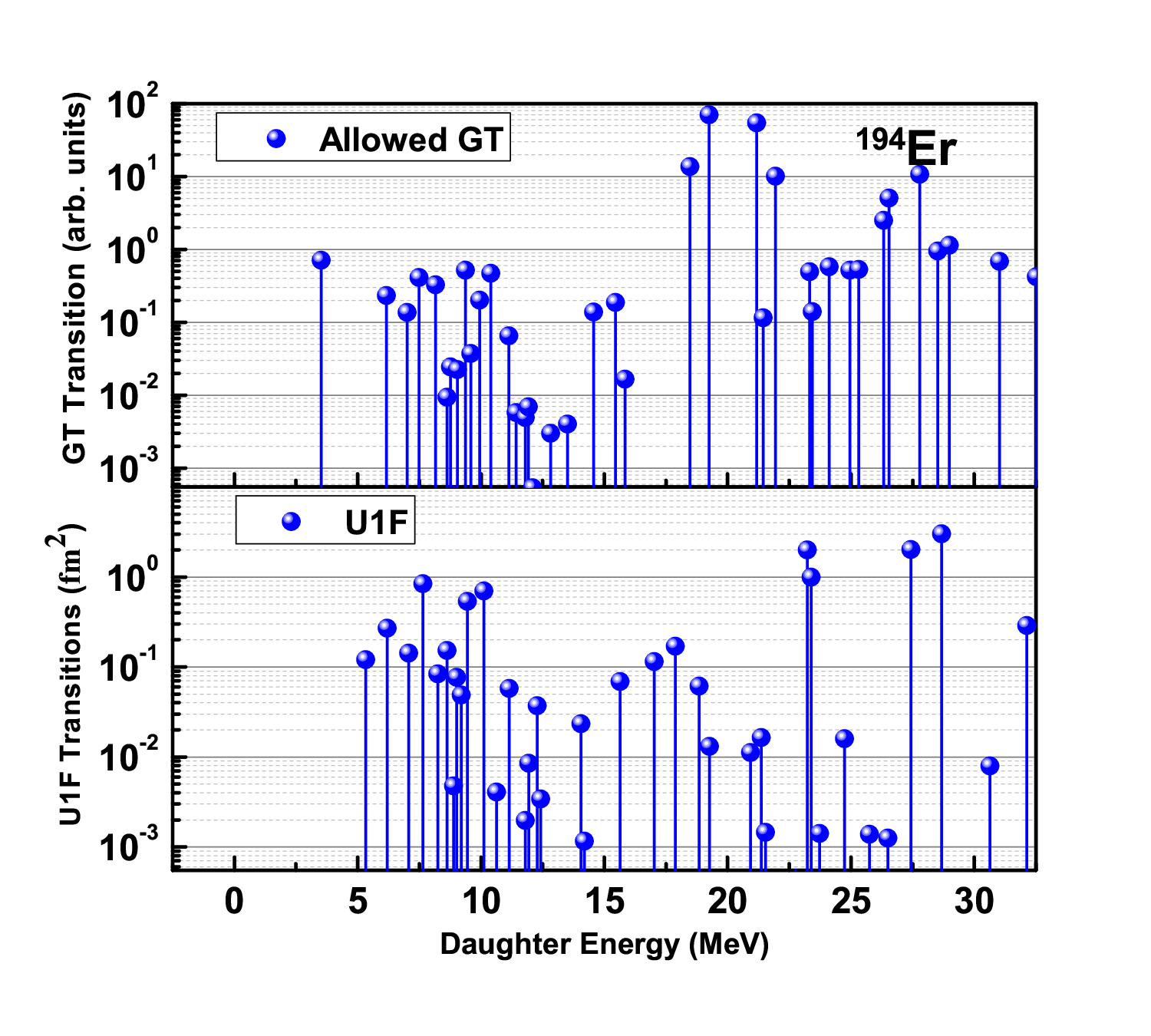}}&
		\resizebox{0.5\hsize}{!}{\includegraphics*{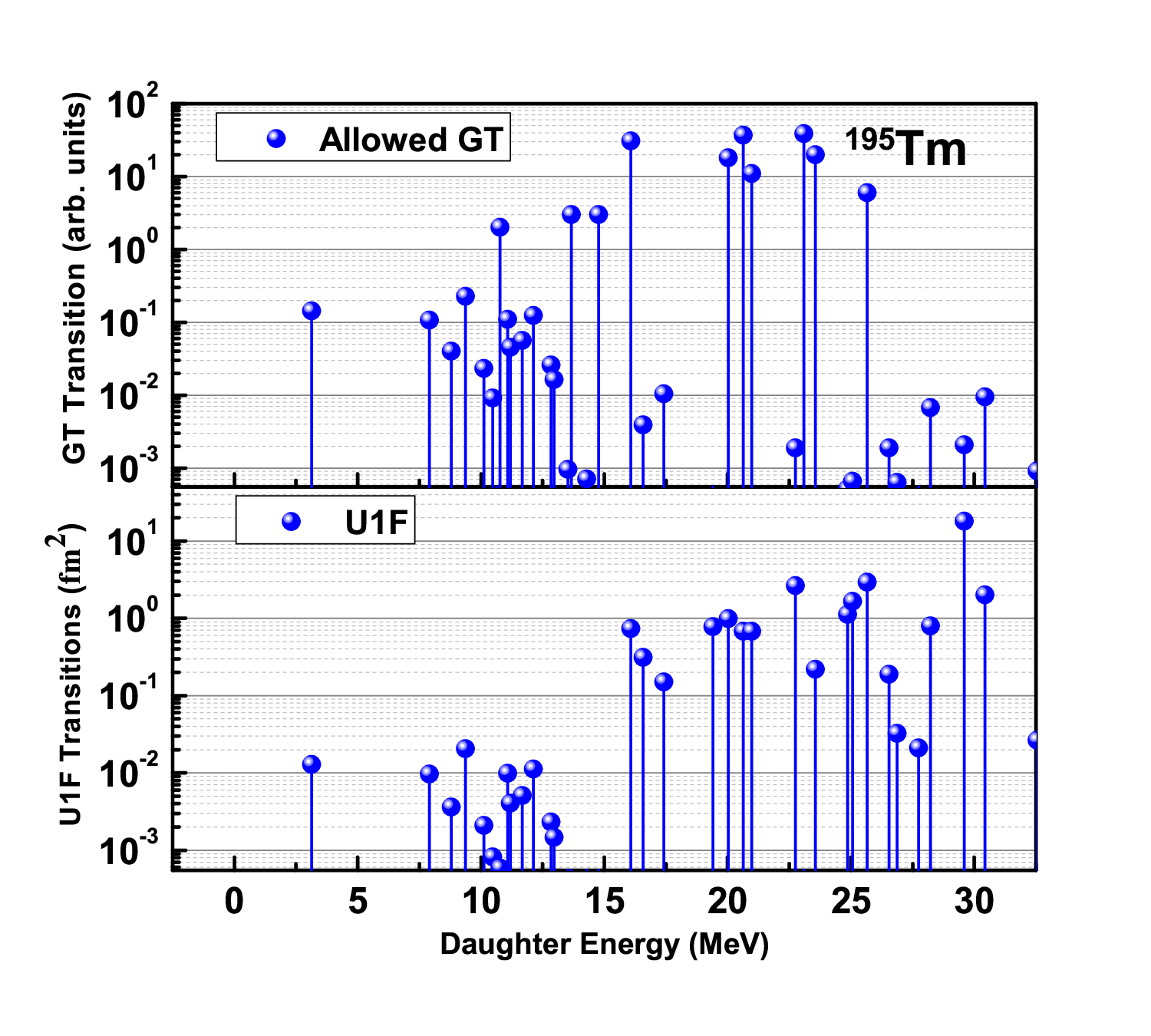}}\\
	\end{tabular}
	\caption{\scriptsize Allowed GT and U1F transitions for $^{190}$Gd, $^{191}$Tb, $^{192}$Dy, $^{193}$Ho,
		$^{194}$Er and $^{195}$Tm as a function excitation energy  in daughter nucleus computed
		using the pn-QRPA (N) model.} \label{fig8} \centering
\end{figure*}
\begin{figure*}
	\begin{tabular}{cc}
		\resizebox{0.5\hsize}{!}{\includegraphics*{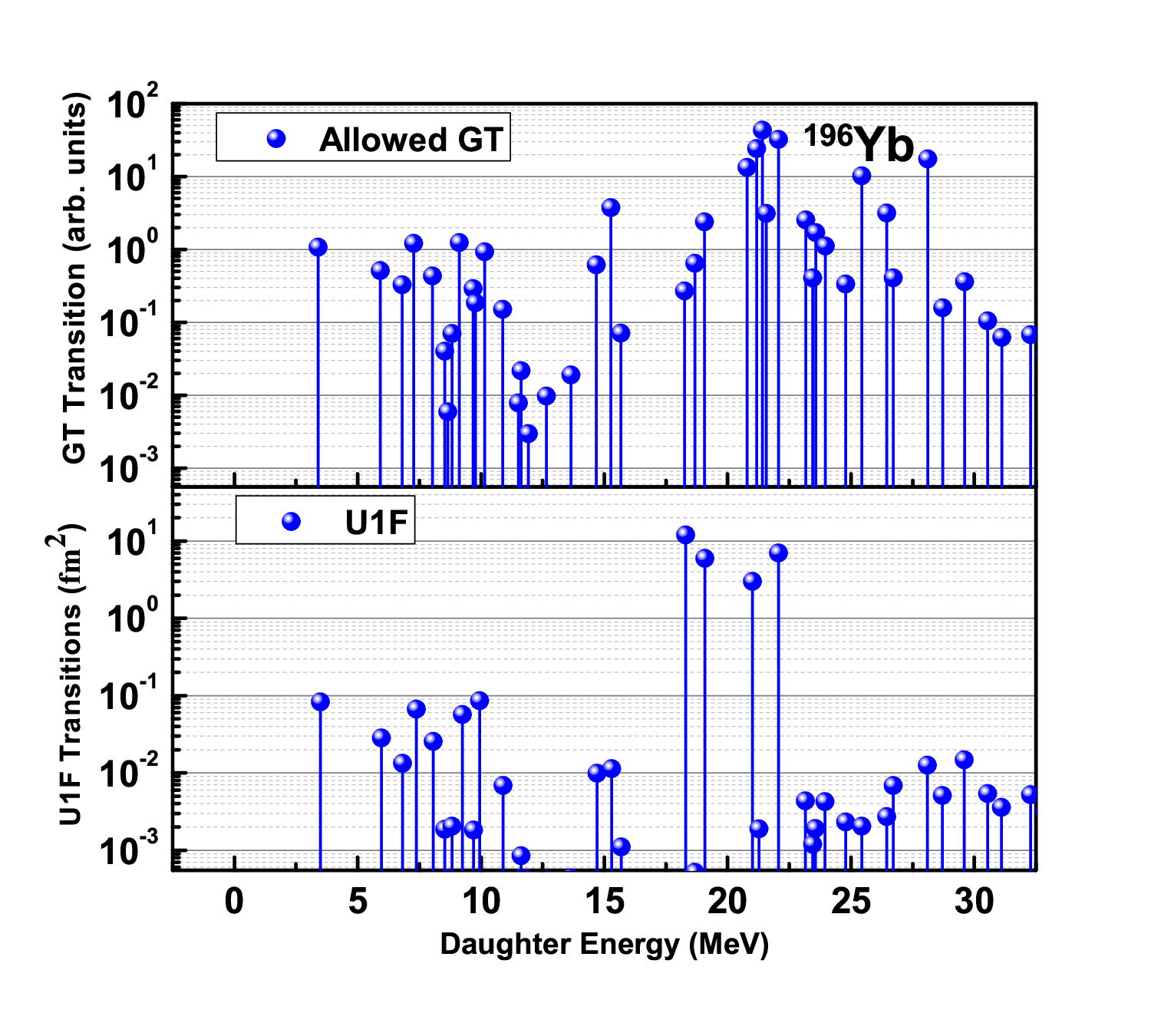}}&
		\resizebox{0.5\hsize}{!}{\includegraphics*{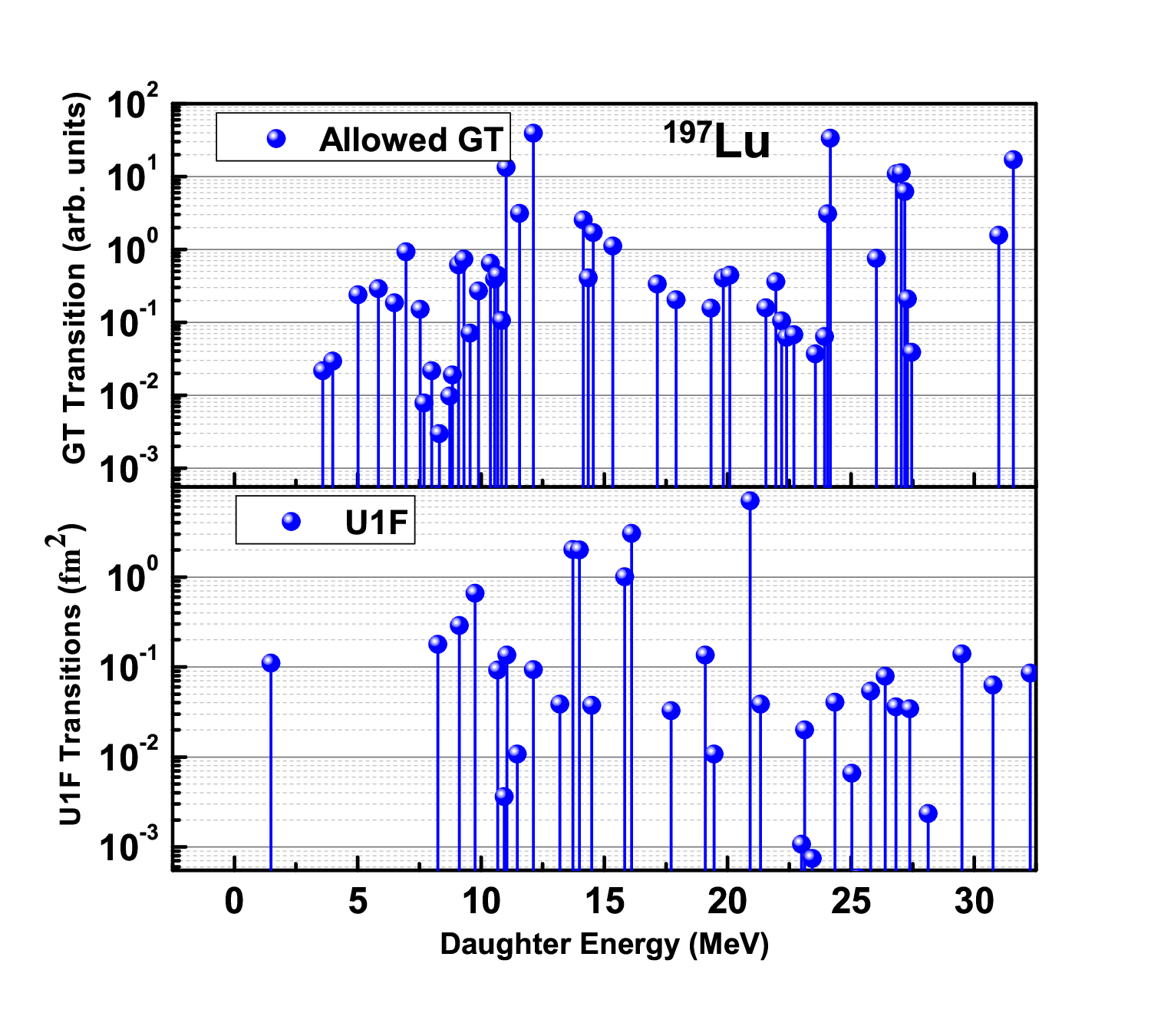}}\\
		\resizebox{0.5\hsize}{!}{\includegraphics*{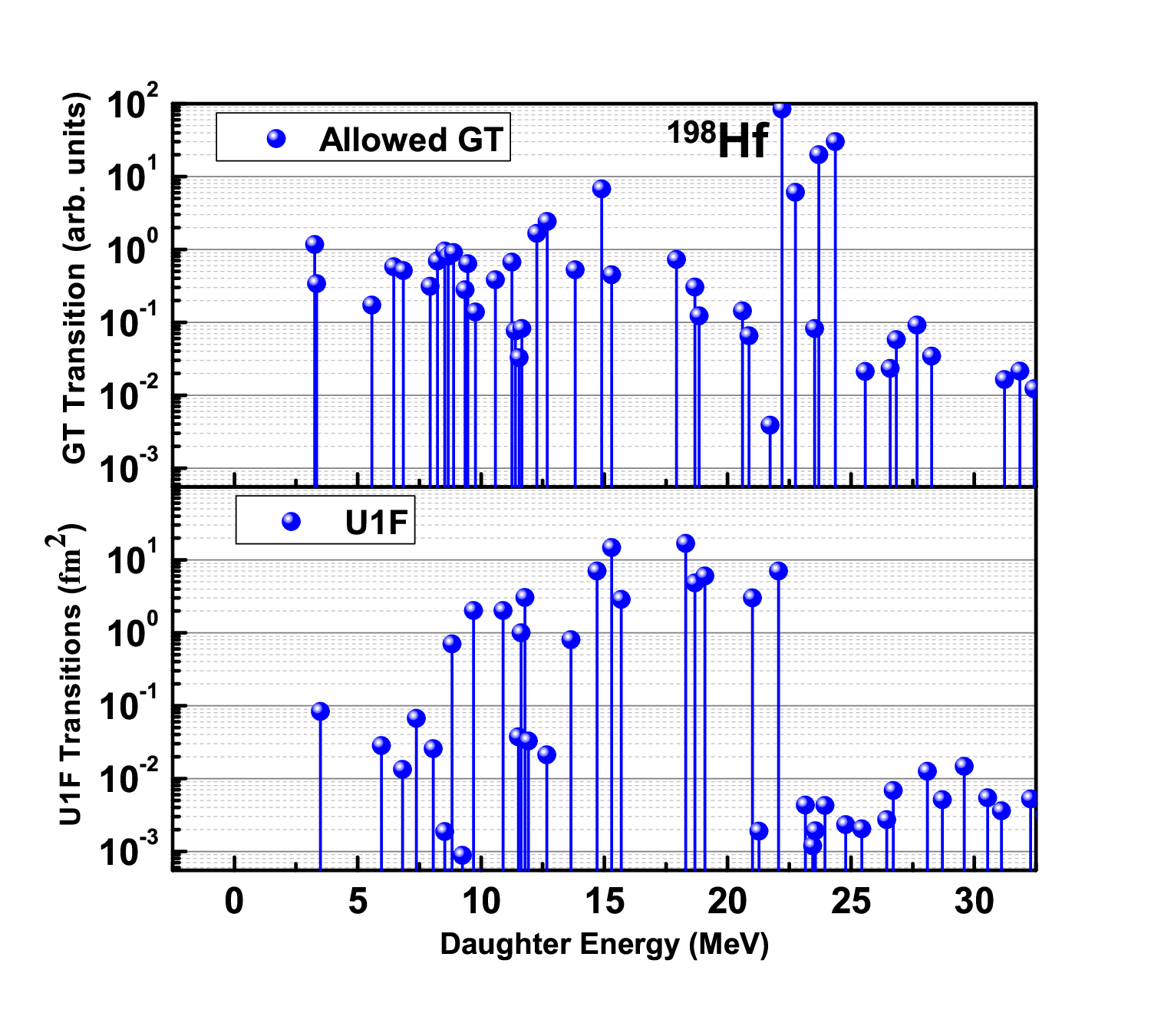}}&
		\resizebox{0.5\hsize}{!}{\includegraphics*{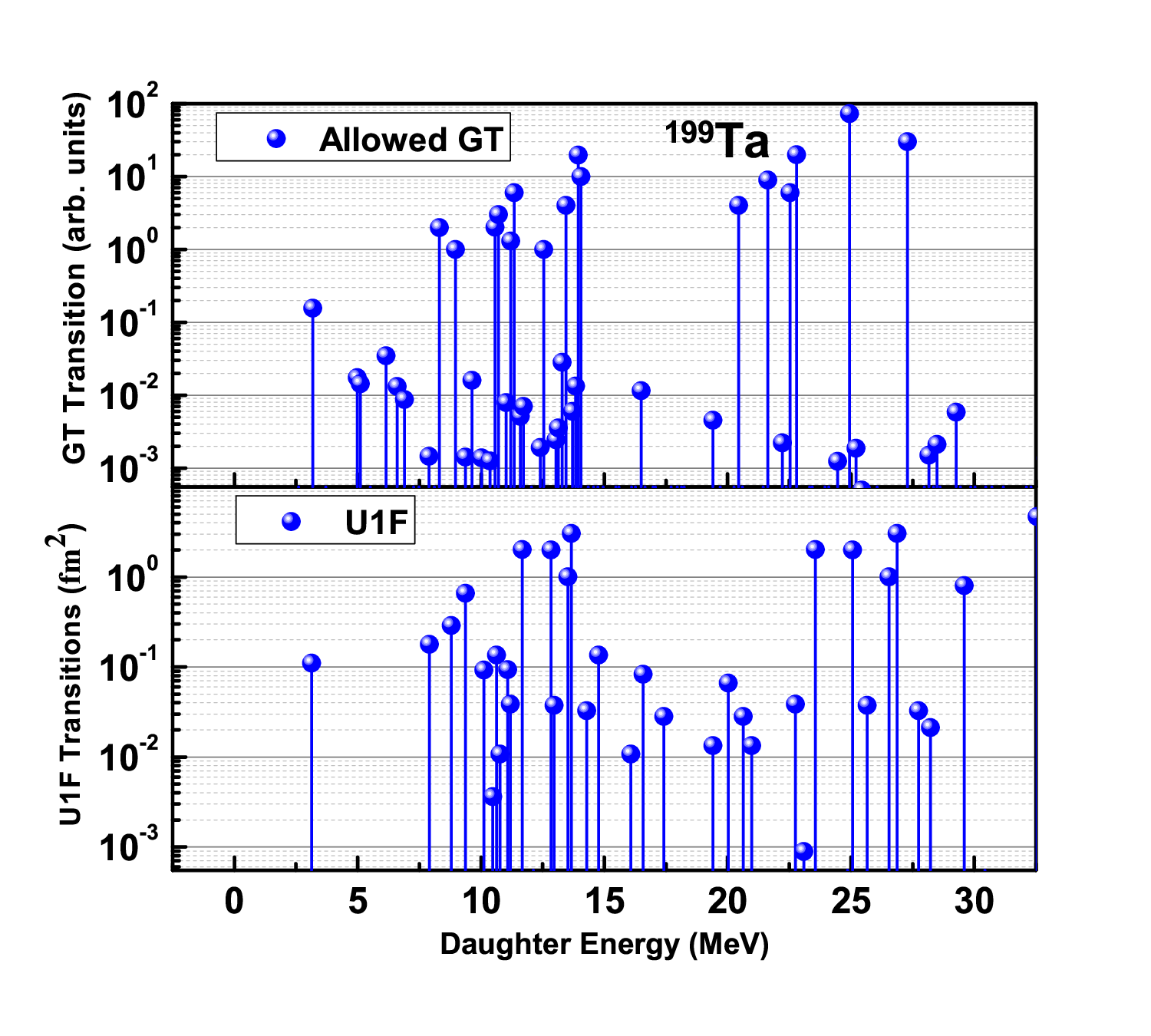}}\\
		\resizebox{0.5\hsize}{!}{\includegraphics*{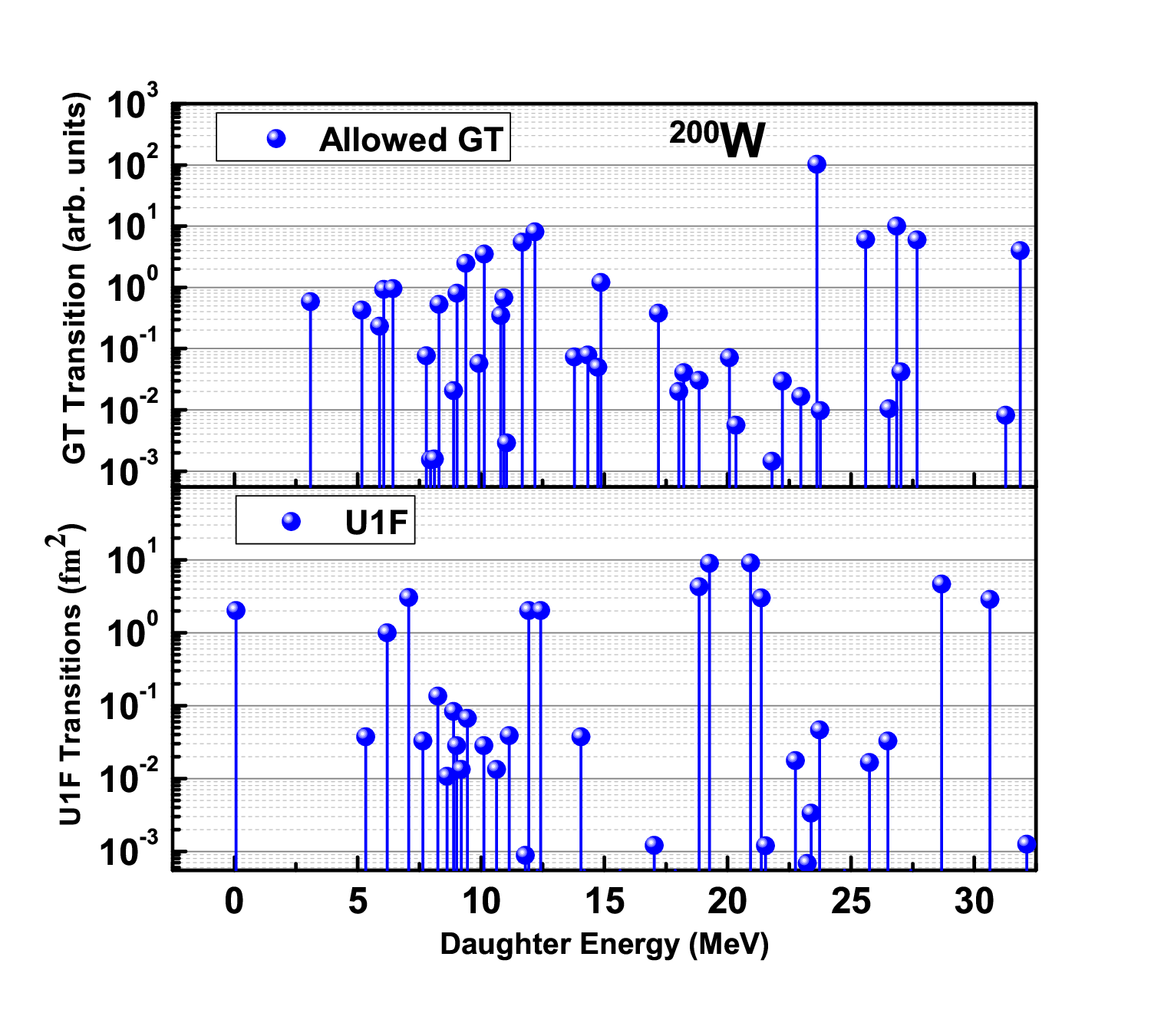}}&
		\resizebox{0.5\hsize}{!}{\includegraphics*{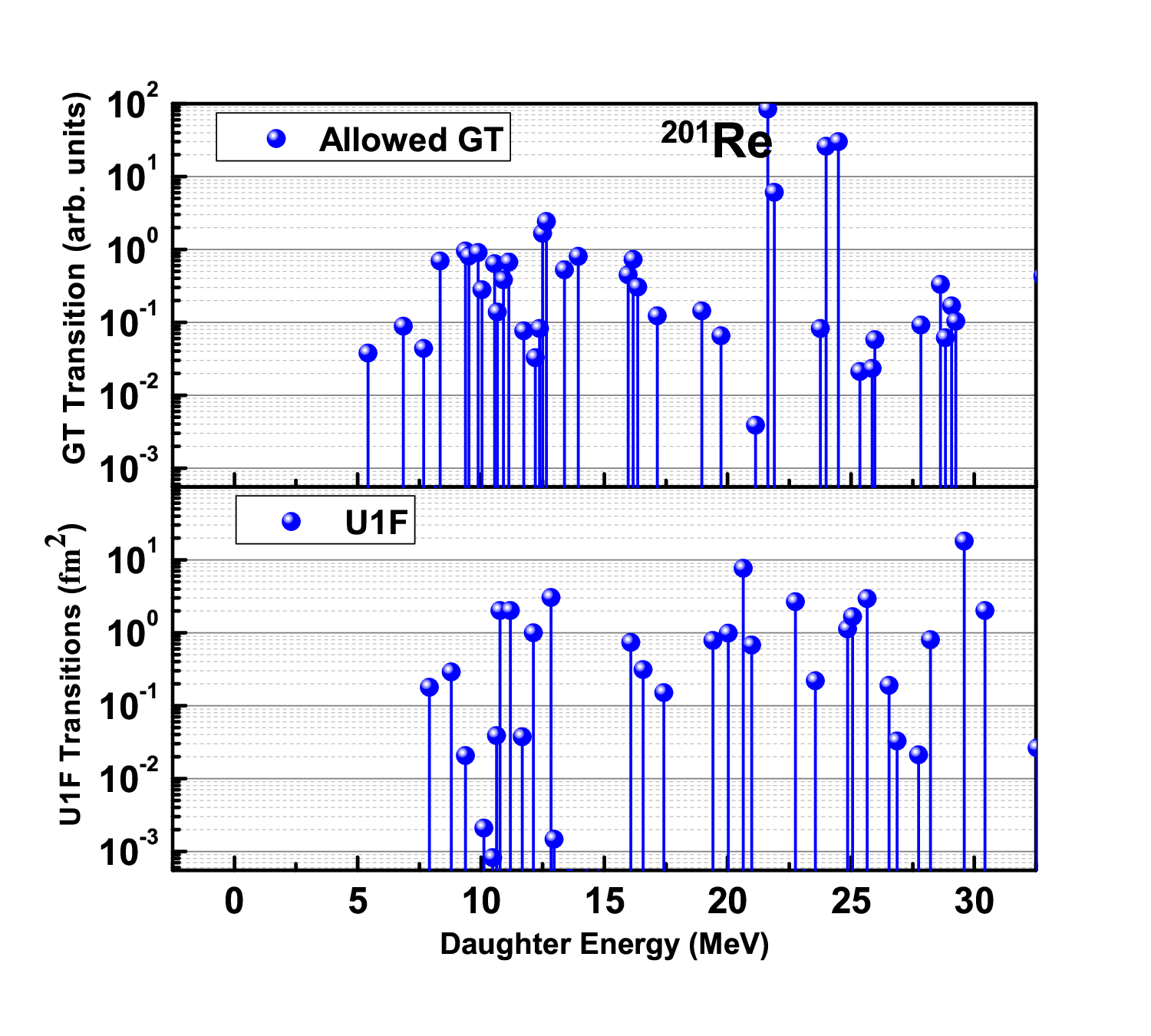}}\\
	\end{tabular}
	\caption{\scriptsize Same as Fig.~\ref{fig8} but for $^{196}$Yb, $^{197}$Lu, $^{198}$Hf,
		$^{199}$Ta, $^{200}$W and $^{201}$Re.}
	\label{fig8b} \centering
\end{figure*}
\begin{figure*}
	\begin{tabular}{cc}
		\resizebox{0.5\hsize}{!}{\includegraphics*{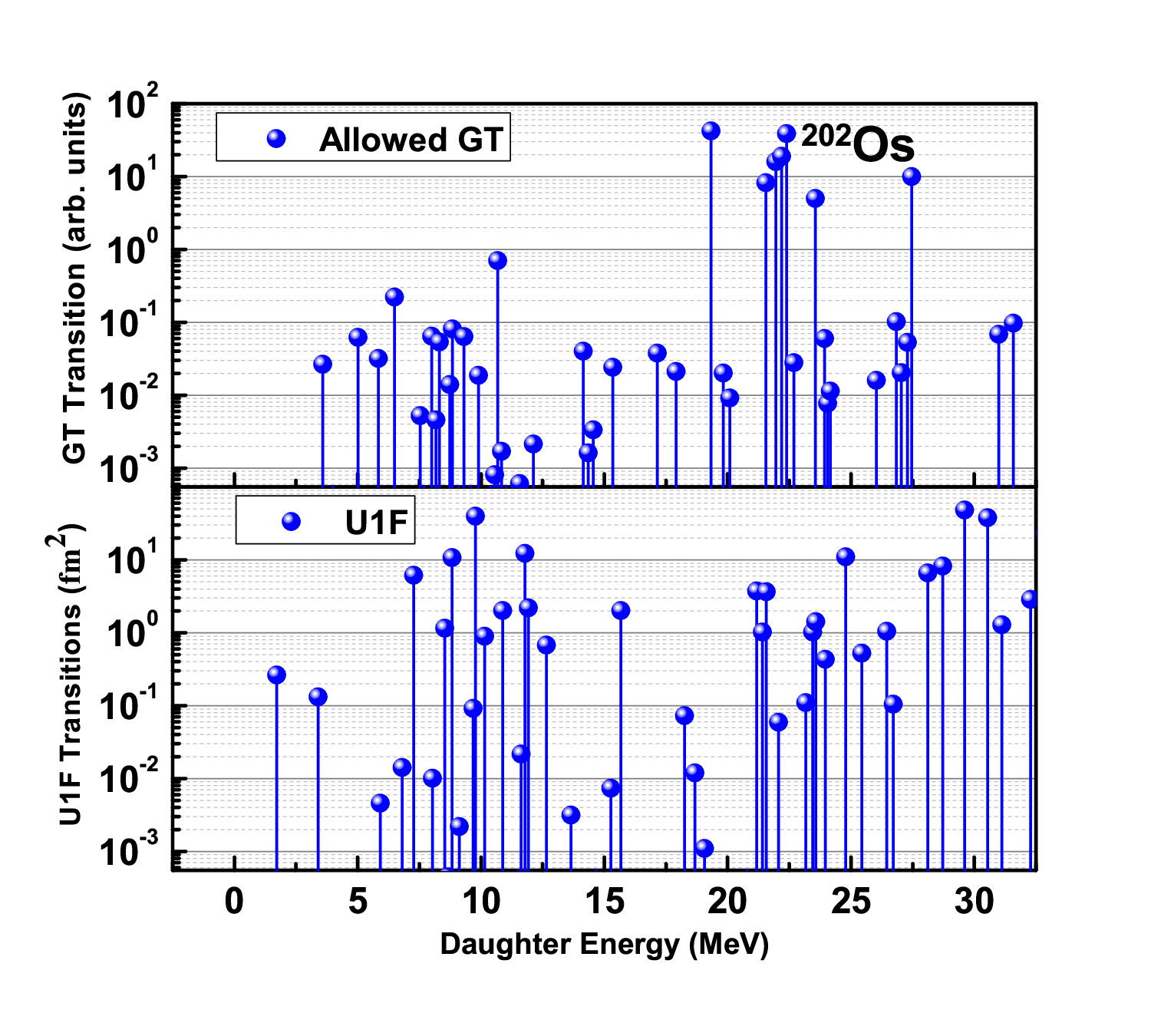}}&
		\resizebox{0.5\hsize}{!}{\includegraphics*{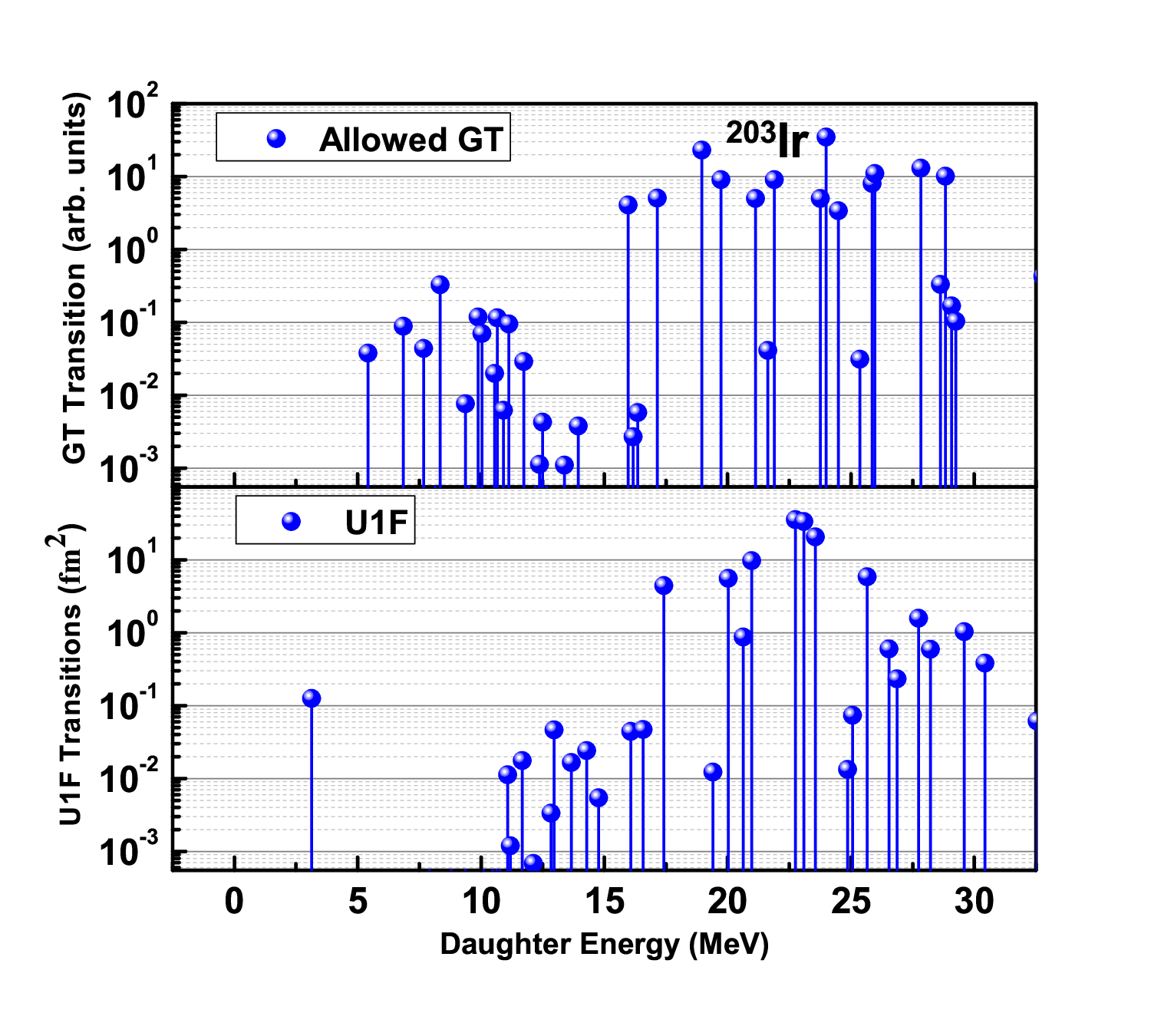}}\\
		\resizebox{0.5\hsize}{!}{\includegraphics*{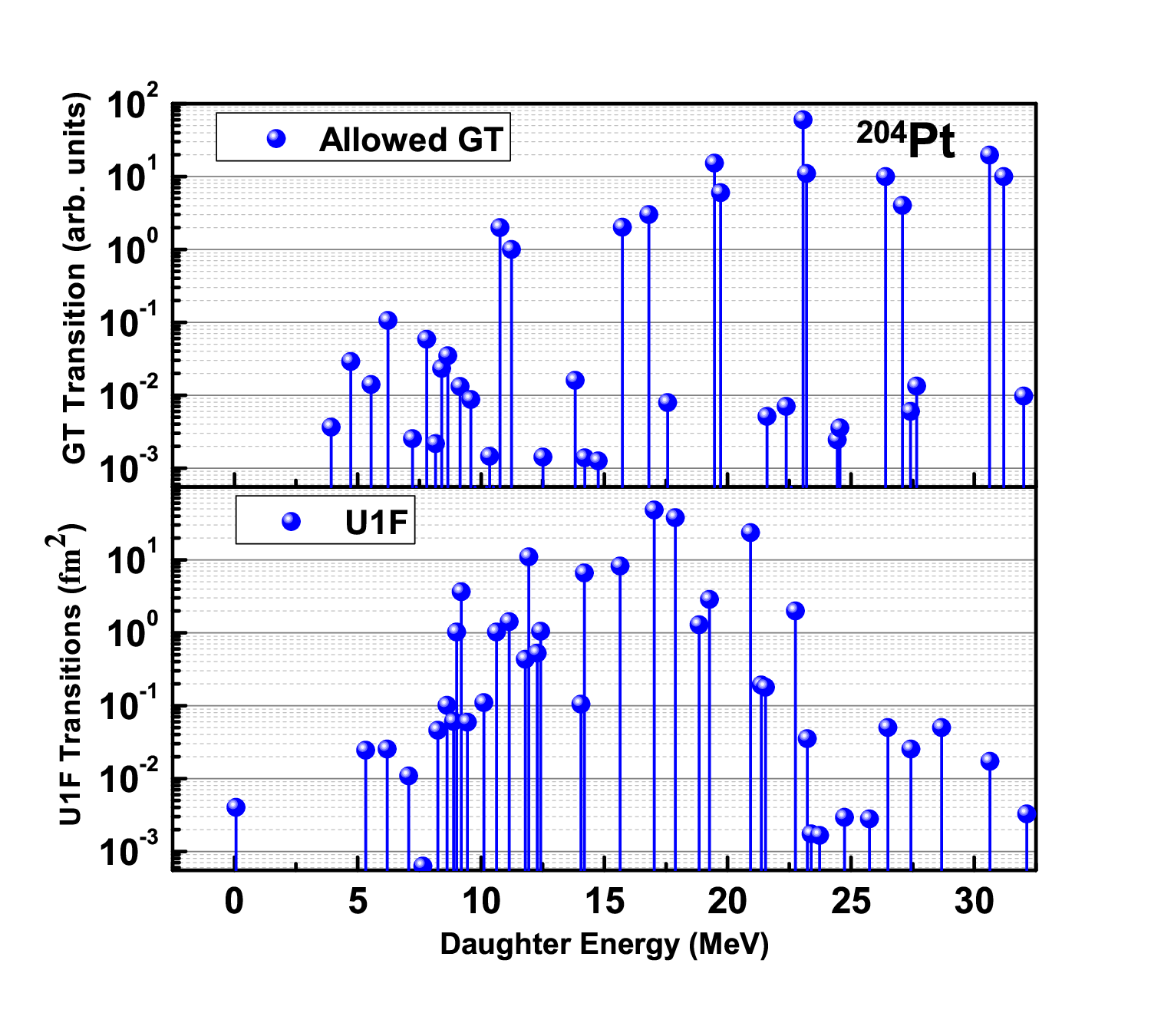}}&
		\resizebox{0.5\hsize}{!}{\includegraphics*{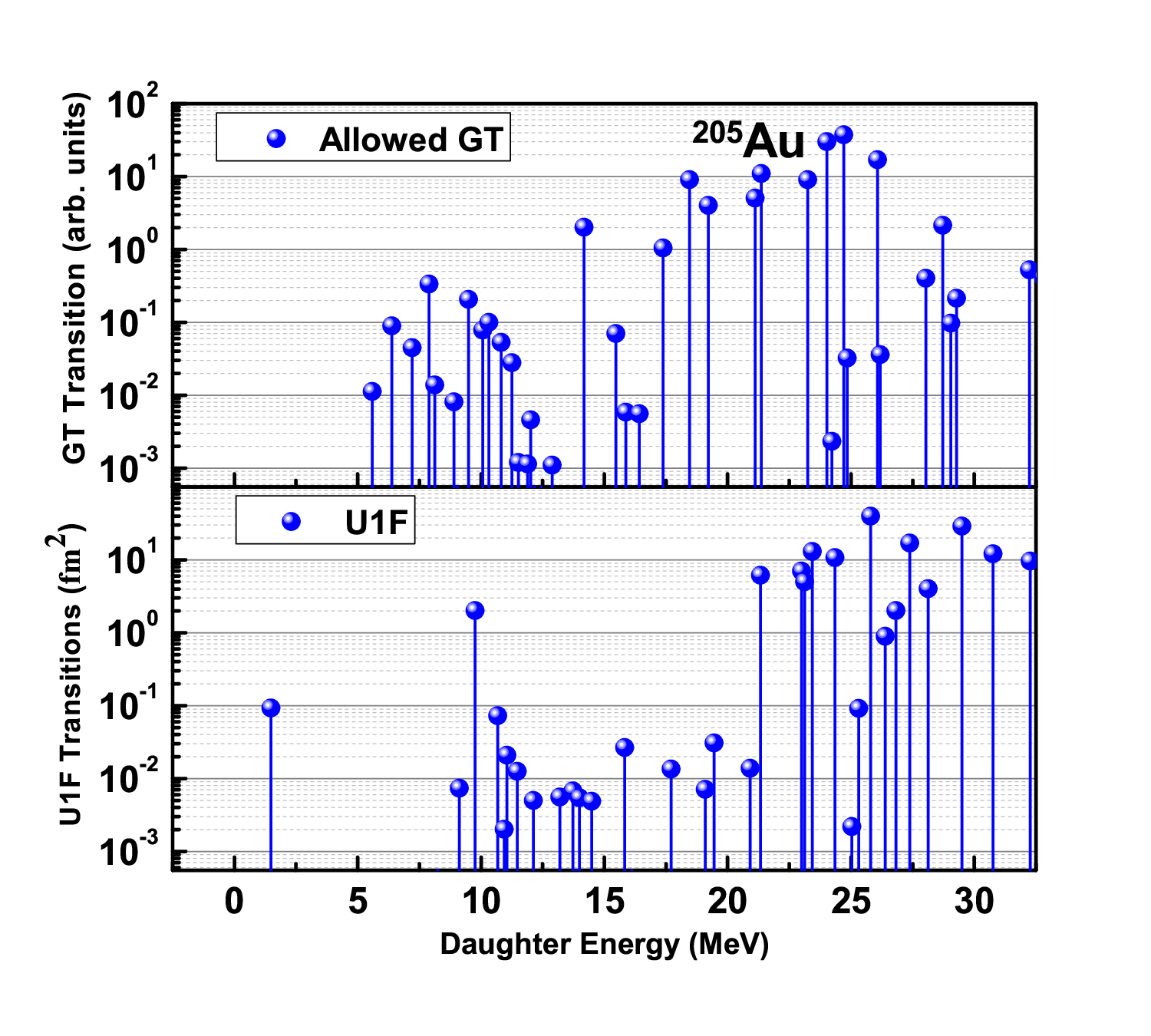}}\\
		\resizebox{0.5\hsize}{!}{\includegraphics*{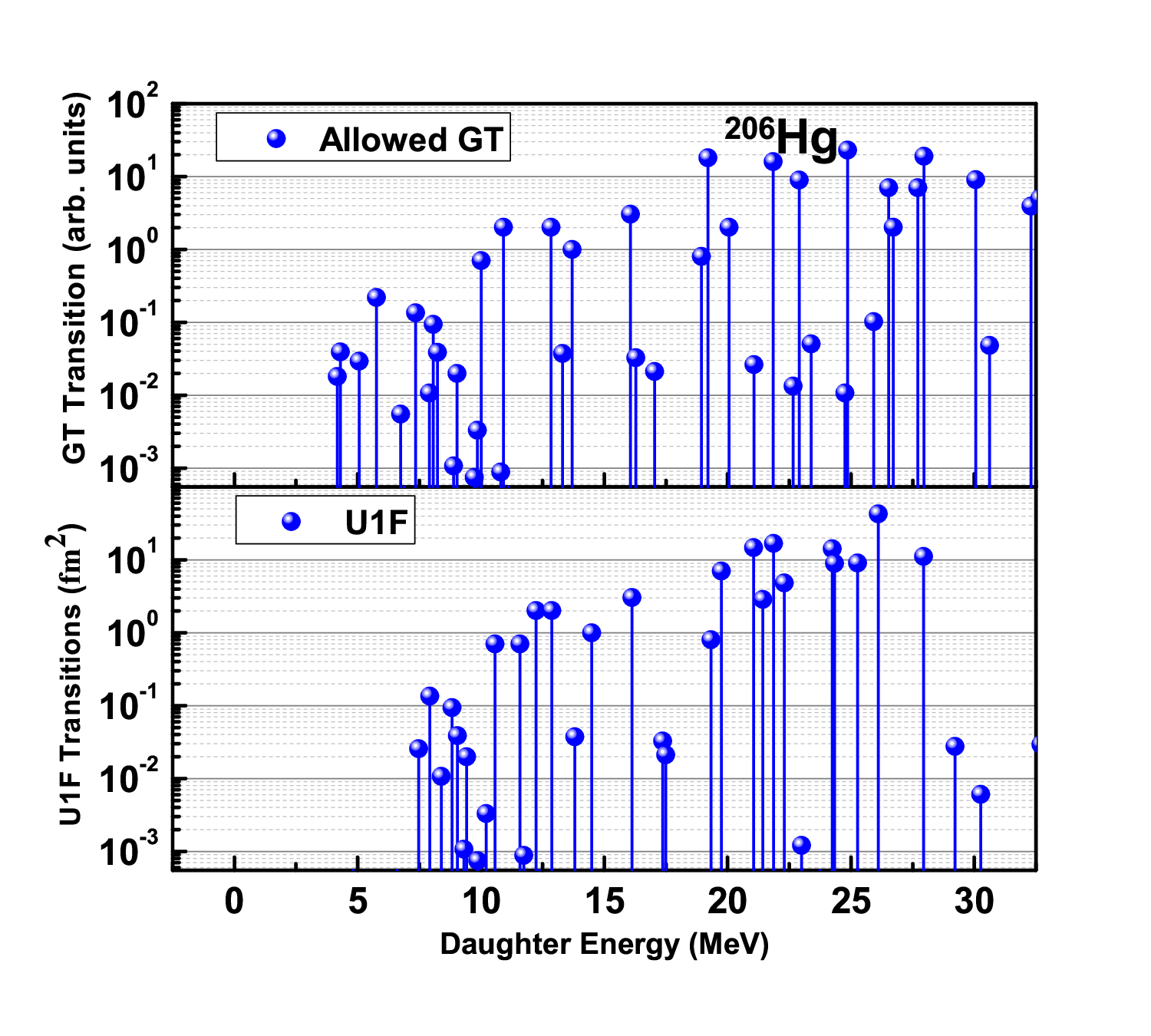}}\\
	\end{tabular}
	\caption{\scriptsize Same as Fig.~\ref{fig8} but for $^{202}$Os, $^{203}$Ir, $^{204}$Pt,
		$^{205}$Au and $^{206}$Hg.}
	\label{fig9} \centering
\end{figure*}

Figs.~(\ref{fig8}-\ref{fig9}) show the pn-QRPA (N) computed GT strength distributions  for the selected $N$ = 126 $r$-process waiting point nuclei. The allowed GT and U1F transitions are shown separately for each nucleus. The allowed GT strengths are given in units such that GT = 3 for neutron decay. We found that U1F transitions contributed significantly in reducing the total $\beta$-decay half-lives, for the cases shown in Fig.~\ref{fig9} whereas the U1F contributions are relatively smaller for the cases displayed in Fig.~\ref{fig8} and Fig.~\ref{fig8b}. The U1F transitions are relatively bigger in magnitude than allowed GT transitions for $^{202}$Os, $^{203}$Ir, $^{204}$Pt and $^{205}$Au. This is in line with the findings of Ref.~\cite{Bor06} that FF contribution dominates over allowed GT  for $Z \geq 76$. Phase space amplification is yet another factor which contributes to the enhancement of U1F transitions. 
The strength distribution depends much on the choice of GT interaction strengths. Usually larger values of \textit{ph} interaction strength shift the GT giant resonance to higher excitation energies. For the cases $^{196}$Yb and $^{198}$Hf, the values of particle-hole interaction strength are smaller as compared to other nuclei. That could be one probable reason why larger U1F strengths are available at lower excitation energies and little beyond 25 MeV.
For $^{206}$Hg, the U1F and allowed GT transitions are comparable in magnitude and the half-life is reduced by $\sim$ 46$\%$ when U1F transition was included (see Table~\ref{tab1}). Our calculation fulfilled the model-independent Ikeda Sum Rule (ISR).

\section{Conclusions}

In this study, we investigated the $\beta$-decay properties of  $N$ = 126 waiting point nuclei using two versions of the pn-QRPA model. Allowed GT and U1F transitions were considered in the pn-QRPA (N) calculation, whereas the pn-QRPA (WS) calculation included allowed GT, rank 0, 1 and 2 excitations. The computed Ikeda sum rule was satisfied for all the cases (up to $1\%$ deviation was noted in few cases of odd-A nuclei). The pn-QRPA computed half-lives were in decent agreement with the experimental data. The incorporation of FF transitions with rank 0 and 1 operator value in the pn-QRPA (N) model may further improve the computed half-lives, which we hope to report in near future. The stellar weak rates were computed as a function of core density and temperature values. These rates were found to increase (decrease) with the increase in stellar temperature (density). This study may contribute to accelerating the $r$-process nucleosynthesis calculation. The shorter half-lives reported in this work as compared to previous calculations may lead to re-adjustment of the third peak of the element abundances toward higher $A$. In the past a similar study of $\beta$-decays of the isotones with $N$ = 126 was performed using shell-model calculations and a modest shift of the third peak of the element abundances in the $r$-process toward a higher mass region was reported \cite{Suz12}. 
We are in a process of computing the nuclear abundances. To date we have performed the necessary  weak rate calculation. We would report further progress after completion of nuclear abundance calculation in near future. 


\vspace{0.5in} \textbf{Acknowledgment}:  
N. \c{C}akmak would like to thank Cevad Selam for very fruitful discussion on calculation of GT and FF transitions.

\newpage

\newpage

\begin{table*}[h!t]%
	\caption{Comparison of $\beta$-decay half-lives of $r$-process nuclei ($N$ = 126) with previous calculations and measured data. } \label{tab1}
	\tiny
	\renewcommand{\arraystretch}{1.0}
	\renewcommand{\tabcolsep}{0.03cm}
	\scriptsize
	\vspace{0.25cm}
	\begin{tabular}{c|ccccccccccccccc}
		\hline
		& \multicolumn{9}{c}{$\beta$-decay half lives [T$_{1/2}$(\textit{s})]} \\
		\cline{2-16}
		\multicolumn{1}{c|}{Nuclei}  & {~~Ref~\cite{Mol97}~~} &
		\multicolumn{3}{c} {~~~~~QRPA-FRDM \cite{Mol03}~~~} &
		\multicolumn{2}{c} {~~~~~SM \cite{Suz12}~~~~~~}&
		\multicolumn{1}{c} {~LSSM \cite{Zhi13}~} &
		\multicolumn{2}{c} {~~CQRPA+DF3 \mbox{\cite{Bor06,Bor11}}~~} &
		\multicolumn{1}{c} {~~Ref~\cite{Zhou17}~~~~}&
		\multicolumn{2}{c} {~~pn-QRPA(WS)~~~~}&
		\multicolumn{2}{c} {~~pn-QRPA(N)~~~~} & {\~Ref~\cite{Aud17}} \vspace{0.2cm}\\
		& (GT) & (Eval) & (GT) & (GT+FF) & (GT) & (GT+FF) & (GT+FF) &(GT) & (GT+FF) & (GT) & (GT) &(GT+FF) &(GT) & (GT+U1F) & \textbf{NUBASE2016} \vspace{0.1cm}\\
		\hline
		
		{\small $^{190}$Gd}   & {\small 0.014}  & {\small 0.016} & {\small 0.014} & {\small 0.010}  & 
		{\small 0.058} & {\small 0.040} & {\small -} & {\small -} & {\small -} & {\small -}  & 
		{\small 0.024} & {\small 0.023} & {\small 0.029} & {\small 0.021} & {\small -} \\
		
		{\small $^{191}$Tb}   & {\small 0.016}  & {\small 0.014} & {\small 0.016} & {\small 0.010}  & 
		{\small 0.077} & {\small 0.053} & {\small -} & {\small 0.019} & {\small 0.005} & {\small -}     &
		{\small 0.061} & {\small 0.050} & {\small 0.020} & {\small 0.016} & {\small -} \\
		
		{\small $^{192}$Dy}   & {\small 0.032}  & {\small 0.030} & {\small 0.032} & {\small 0.020}  & 
		{\small 0.012} & {\small 0.078} & {\small 0.010} & {\small 0.029} & {\small 0.005} & {\small 0.021} & 
		{\small 0.045} & {\small 0.031} & {\small 0.049} & {\small 0.035} & {\small -} \\
		
		{\small $^{193}$Ho}   & {\small 0.028}  & {\small 0.021} & {\small 0.028} & {\small 0.018}  & 
		{\small 0.017} & {\small 0.011} & {\small 0.014} & {\small 0.008} & {\small 0.017} & {\small 0.028} &
		{\small 0.016} & {\small 0.012} & {\small 0.022} & {\small 0.021} & {\small -} \\
		
		{\small $^{194}$Er}   & {\small 0.087}  & {\small 0.096} & {\small 0.087} & {\small 0.051}  & 
		{\small 0.029} & {\small 0.081} & {\small 0.025} & {\small 0.012} & {\small 0.029} & {\small 0.036} &
		{\small 0.094} & {\small 0.078} & {\small 0.171} & {\small 0.095} & {\small -} \\
		
		{\small $^{195}$Tm}   & {\small 0.067}  & {\small 0.091} & {\small 0.067} & {\small 0.042}  & 
		{\small 0.055} & {\small 0.029} & {\small 0.036} & {\small 0.016} & {\small 0.055} & {\small 0.049} &
		{\small 0.044} & {\small 0.038} & {\small 0.118} & {\small 0.037} & {\small -} \\
		
		{\small $^{196}$Yb}   & {\small 0.397}  & {\small 0.222} & {\small 0.397} & {\small 0.181}  & 
		{\small 0.103} & {\small 0.044} & {\small 0.069} & {\small 0.023} & {\small 0.102} & {\small 0.067} &
		{\small 0.371} & {\small 0.203} & {\small 0.330} & {\small 0.177} & {\small -} \\
		
		{\small $^{197}$Lu}   & {\small 0.119}  & {\small -} & {\small -} & {\small -} & 
		{\small 0.223}  & {\small 0.085} & {\small 0.108} & {\small -}  & {\small 0.079} & {\small 0.094} & 
		{\small 0.415} & {\small 0.182} & {\small 0.370} & {\small 0.259} & {\small -} \\
		
		{\small $^{198}$Hf}   & {\small 3.156}  & {\small -} & {\small -} & {\small -} & 
		{\small 0.505}  & {\small 0.130} & {\small 0.193} & {\small -}  & {\small 0.228} & {\small 0.136} & 
		{\small 2.751} & {\small 1.377} & {\small 1.905} & {\small 1.618} & {\small -} \\
		
		{\small $^{199}$Ta}   & {\small 0.701}  & {\small -} & {\small -} & {\small -} & 
		{\small 1.584}  & {\small 0.279} & {\small 0.286} & {\small -}  & {\small 0.374} & {\small 0.205} & 
		{\small 1.372} & {\small 0.618} & {\small 0.798} & {\small 0.762} & {\small -} \\
		
		{\small $^{200}$W}   & {\small $>$100}  & {\small -} & {\small -} & {\small -} & 
		{\small -}  & {\small -} & {\small -} & {\small -}  & {\small -} & {\small -} & 
		{\small 14.285} & {\small 6.812} & {\small 11.717} & {\small 8.764} & {\small -} \\
		
		{\small $^{201}$Re}   & {\small 1.190}  & {\small -} & {\small -} & {\small -} & 
		{\small -}  & {\small -} & {\small -} & {\small -}  & {\small -} & {\small -} & 
		{\small 2.374} & {\small 1.029} & {\small 2.157} & {\small 1.438} & {\small -} \\
		
		{\small $^{202}$Os}   & {\small $>$100}  & {\small -}  & {\small -} & {\small -} & 
		{\small -}  & {\small -} & {\small -} & {\small -}  & {\small -} & 
		{\small -}  & {\small 8.572} & {\small 0.186}  & {\small 7.620} & {\small 0.207}& {\small 0.2} \\
		
		{\small $^{203}$Ir}   & {\small $>$100}  & {\small -}  & {\small -} & {\small -} & 
		{\small -}  & {\small -} & {\small -} & {\small -}  & {\small -} & 
		{\small -}  & {\small 10.932} & {\small 5.071} & {\small 28.237}  & {\small 7.110} & {\small 6.0} \\
		
		{\small $^{204}$Pt}   & {\small $>$100}  & {\small -}  & {\small -} & {\small -} & 
		{\small -}  & {\small -} & {\small -} & {\small -}  & {\small -} & 
		{\small -}  & {\small 72.003} & {\small 8.716} & {\small 65.588}  & {\small 13.659} & {\small 10.3} \\
		
		{\small $^{205}$Au}   & {\small $>$100}  & {\small -}  & {\small -} & {\small -} & 
		{\small -}  & {\small -} & {\small -} & {\small -}  & {\small -} & 
		{\small -}  & {\small 94.173} & {\small 29.033} & {\small 138.301}  & {\small 37.843} & {\small 32.5} \\
		
		{\small $^{206}$Hg}   & {\small $>$100}  & {\small -}  & {\small -} & {\small -} & 
		{\small -}  & {\small -} & {\small -} & {\small -}  & {\small -} & 
		{\small -}  & {\small 1071.325} & {\small 387.395} & {\small 1116.414}  & {\small 606.287} & {\small 499.2} \\
		\hline
	\end{tabular}
\end{table*}

\clearpage
\begin{table*}
	\centering \caption{Calculated $\beta$-decay rates (allowed GT and U1F) of $N$ = 126 isotones for different stellar temperatures (in units of $GK$) and densities $\rho$Y$_{e}$ (in units of \textit{gcm$^{-3}$}). The rates are tabulated in log to base 10 values (in units of \textit{s$^{-1}$}).} \label{tab2}
	\centering \footnotesize\setlength{\tabcolsep}{1.5pt}
	\renewcommand{\arraystretch}{1.0}
	\renewcommand{\tabcolsep}{0.15cm}
	\vspace{0.05cm}
	\begin{tabular}{|c|c|c|c|c|c|c|c|}
		
		\hline
		Nuclei & $T_{9}$ & \multicolumn{3}{c|}{$\lambda^{\beta}$(\textit{s$^{-1}$)}(Allowed GT)} &
		\multicolumn{3}{c|}{$\lambda^{\beta}$(\textit{s$^{-1}$)}(U1F)} \\
		\cline{3-8} & & $\rho$$\it Y_{e}$=10$^{2}$ & $\rho$$\it
		Y_{e}$=10$^{6}$ & $\rho$$\it Y_{e}$=10$^{10}$ & $\rho$$\it
		Y_{e}$=10$^{2}$ & $\rho$$\it Y_{e}$=10$^{6}$ &
		$\rho$$\it Y_{e}$=10$^{10}$\\
		\hline
		&1  & 1.513  & 1.509  & -16.78 & 0.912  & 0.910  & -7.041  \\
		&2  & 1.513  & 1.510  & -8.343 & 0.979  & 0.978  & -1.421  \\
		$^{190}$Gd  &5  & 1.596  & 1.595  & -2.606 & 3.522  & 3.522  & 2.102  \\
		&15 & 2.404  & 2.404  & 0.640 & 4.795  & 4.795  & 3.894  \\
		&25 & 2.484  & 2.484  & 1.475 & 4.807  & 4.807  & 4.230  \\
		\hline
		&1  & 1.687  & 1.684  & -10.48 & 0.996  & 0.995  & -20.90  \\
		&2  & 1.961  & 1.959  & -6.332 & 0.971  & 0.971  & -11.07  \\
		$^{191}$Tb  &5  & 2.288  & 2.287  & -1.764 & 1.013  & 1.013  & -4.528  \\
		&15 & 3.148  & 3.148 & 1.463 & 1.928  & 1.928  & 0.019  \\
		&25 & 3.247  & 3.247 & 2.284 & 2.086  & 2.086  & 1.074  \\
		\hline
		&1 & 1.239	&1.234&	-22.421 &0.565	&0.561	&-14.91\\
		&2 & 1.239	&1.235&	-11.123	&0.609	&0.606	&-6.272\\
		$^{192}$Dy	&5 & 1.323	&1.322&	-3.587	&2.947	&2.947	&-0.298\\
		&15& 2.465	&2.465&	0.756	&4.500	&4.500	&3.096\\
		&25& 2.818	&2.818&	1.859	&4.606	&4.606	&3.809\\
		\hline
		&1	&1.442	&1.439	&-14.75 &1.748	&1.748&	-13.46\\
		&2	&1.402	&1.400	&-8.095	&2.819	&2.818&	-5.841\\
		$^{193}$Ho    	&5	&1.599	&1.599	&-2.853	&3.436	&3.436&	-0.141\\
		&15	&2.870	&2.870	&1.140	&4.518	&4.518&	3.268\\
		&25	&3.033	&3.033	&2.070	&4.749	&4.749&	4.032\\
		\hline
		&1&	0.795&	0.788&	-26.62&-0.038&	-0.045&	-17.15\\
		&2&	0.794&	0.790&	-13.15	&0.064&	0.06&	-7.413\\
		$^{194}$Er	&5&	0.933&	0.932&	-4.27&	2.768&	2.768&	-0.837\\
		&15&2.174&	2.174&	0.462&	4.262&	4.262&	2.773\\
		&25&2.529&	2.529&	1.582&	4.351&	4.35&	3.518\\
		\hline
		&1	&0.633&	0.629&	 -21.50& 0.391&	 0.391&	-21.35\\
		&2	&0.658&	0.655&	 -11.66& 0.343&	 0.342&	-11.58\\
		$^{195}$Tm 	&5	&1.162&	1.161&	 -3.801& 0.170&	 0.170&	-5.278\\
		&15	&2.927&	2.927&	  1.033&-0.143&	-0.143&	-2.061\\
		&25	&3.145&	3.145&	  2.110&-0.385&	-0.385&	-1.402\\
		\hline
		&1&	0.397&	0.384&	-27.58&	-0.251&	-0.254&	-37.99\\
		&2&	0.397&	0.387&	-14.06&	-0.251&	-0.253&	-19.83\\
		$^{196}$Yb 	&5&	0.651&	0.649&	-5.121&	-0.241&	-0.242&	-8.425\\
		&15&2.046&	2.046&	0.301&	0.516&	0.515&	-2.086\\
		&25&2.445&	2.445&	1.546&	0.661&	0.661&	-0.674\\
		\hline
		&1  & 0.262  & 0.249  & -27.640 & -1.910  & -1.929 & -37.810  \\
		&2  & 0.261  & 0.252  & -14.140& -0.467 & -0.469  & -19.070  \\
		$^{197}$Lu  &5  & 0.509  & 0.507  & -5.211 & 0.517  & 0.516  &  -7.331 \\
		&15 & 1.967  & 1.966  &  0.234&  1.516 & 1.516  & -1.019  \\
		&25 & 2.391  & 2.391  &  1.496 & 1.717  & 1.717  & 0.414  \\
		\hline
		&1  & -0.439  & -0.498  & -32.100 & -1.190  & -1.219  & -48.400  \\
		&2  & -0.441 &  -0.480  & -17.070 & -1.191  & -1.213  & -25.18  \\
		$^{198}$Hf  &5  &  0.240  &  0.237 & -7.036 & -1.189  & -1.195  & -10.87  \\
		&15 &  1.476 &  1.476 & -0.997 & -0.474 & -0.475 & -3.451  \\
		&25 &  1.514 &  1.513 & 0.235 & -0.392  & -0.392  & -1.907  \\
		\hline
	\end{tabular}
\end{table*}

\clearpage
\begin{table*}
	\centering \caption{Same as Table~\ref{tab2} but for heavier nuclei.} \label{tab3}
	\centering \footnotesize\setlength{\tabcolsep}{1.5pt}
	\renewcommand{\arraystretch}{1.1}
	\renewcommand{\tabcolsep}{0.15cm}
	\vspace{0.2cm}
	\begin{tabular}{c|c|c|c|c|c|c|c|}
		
		\hline
		Nuclei & $T_{9}$ & \multicolumn{3}{c|}{$\lambda^{\beta}$(\textit{s$^{-1}$})(Allowed GT)} & 
		\multicolumn{3}{c|}{$\lambda^{\beta}$(\textit{s$^{-1}$)}(U1F)} \\
		\cline{3-8} & & $\rho$$\it Y_{e}$=10$^{2}$ & $\rho$$\it
		Y_{e}$=10$^{6}$ & $\rho$$\it Y_{e}$=10$^{10}$ & $\rho$$\it
		Y_{e}$=10$^{2}$ & $\rho$$\it Y_{e}$=10$^{6}$ &
		$\rho$$\it Y_{e}$=10$^{10}$\\
		\hline
		&1  & -0.119 & -0.128 & -26.675 & -1.771 & -1.843 & -34.360 \\
		&2  & -0.143 & -0.150  & -14.640  & -1.763 & -1.809 & -18.600  \\
		$^{199}$Ta	&5  & 0.880   & 0.879  & -5.734  & -0.099 & -0.102 & -8.115 \\
		&15 & 2.265  & 2.264  & 0.206   & 1.097  & 1.096  & -1.460  \\
		&25 & 2.483  & 2.483  & 1.403   & 1.122  & 1.122  & -0.188\\
		\hline
		& 1  & -1.228 & -1.348 & -36.587 & -2.281 & -2.374 & -52.431 \\
		& 2  & -1.231 & -1.303 & -19.460  & -2.285 & -2.345 & -27.586 \\
		$^{200}$W& 5  & -0.130  & -0.134 & -8.076  & -2.278 & -2.289 & -12.369 \\
		& 15 & 1.206  & 1.205  & -1.372  & -1.318 & -1.319 & -4.377  \\
		& 25 & 1.243  & 1.243  & -0.078  & -1.217 & -1.217 & -2.773 \\
		\hline
		& 1  & -0.491 & -0.523 & -29.244 & -0.705 & -0.739 & -45.322 \\
		& 2  & -0.492 & -0.515 & -15.540  & -0.707 & -0.730  & -23.635 \\
		$^{201}$Re& 5  & -0.066 & -0.069 & -6.260   & -0.716 & -0.721 & -10.120  \\
		& 15 & 1.569  & 1.569  & -0.230   & 0.014  & 0.013  & -2.900    \\
		& 25 & 2.066  & 2.066  & 1.156   & 0.162  & 0.161  & -1.326 \\
		\hline
		&1&	-1.041&	-1.075&	-38.96&	0.513&	0.505&	-38.92\\
		&2&	-1.042&	-1.065&	-20.13&	0.512&	0.506&	-20.00\\
		$^{202}$Os	&5&	-0.629&	-0.633&	-7.973&	0.722&	0.720&	-7.815\\
		&15&	0.589&	0.589	&-1.745&	1.825&	1.825&	-0.877\\
		&25&	0.878&	0.877&	-0.287&	2.009&	2.009&	0.624\\
		\hline
		&1&	-1.426&	-1.458&	-35.41&	-1.151&	-1.184&	-35.22\\
		&2&	-1.496&	-1.519&	-19.45&	-1.219&	-1.243&	-19.26\\
		$^{203}$Ir 	&5&	-1.475&	-1.480&	-9.122&	-1.200&	-1.205&	-8.927\\
		&15&	-0.071	&-0.071&	-2.722&	0.189&	0.189&	-2.483\\
		&25&	0.211&	0.211&	-1.170&	0.469&	0.469&	-0.921\\
		\hline
		&1&	-2.099&	-2.163&	-44.56&	-1.247&	-1.274&	-46.02\\
		&2&	-2.101&	-2.142&	-23.14&	-1.249&	-1.268&	-24.31\\
		$^{204}$Pt	&5&	-1.195&	-1.199&	-9.181&	-1.013&	-1.018&	-10.38\\
		&15&	0.327&	0.327&	-2.153&	0.176&	0.175&	-2.727\\
		&25&	0.666&	0.666&	-0.558&	0.353&	0.353&	-1.128\\
		\hline
		&1&	-2.270&	-2.330&	-41.10&	-1.986&	-2.047&	-40.88\\
		&2&	-2.403&	-2.442&	-22.39&	-2.120&	-2.159&	-22.17\\
		$^{205}$Au 	&5&	-2.094&	-2.100&	-10.42&	-1.814&	-1.821&	-10.19\\
		&15&-0.520&	-0.520&	-3.277&	-0.251&	-0.252&	-3.020\\
		&25&-0.243&	-0.243&	-1.663&	0.023&	0.023&	-1.401\\
		\hline
		&1&	-3.156&	-3.356&	-50.77&	-3.167&	-3.319&	-53.90\\
		&2&	-3.158&	-3.263&	-26.16&	-3.173&	-3.260&	-28.79\\
		$^{206}$Hg	&5&	-1.821&	-1.826&	-10.49&	-2.804&	-2.817&	-12.94\\
		&15&-0.190&	-0.190&	-2.651&	-1.628&	-1.629&	-4.706\\
		&25&0.287&	0.287&	-0.906&	-1.421&	-1.421&	-2.987\\
		\hline
		
	\end{tabular}
\end{table*}


\begin{table*}
	\centering \caption{Comparison between theoretical calculations of  $\beta$-delayed neutron emission probability values
		for waiting point nuclei.} \label{tab6}
	\small
	\footnotesize\setlength{\tabcolsep}{1.5pt}
	\renewcommand{\arraystretch}{1.1}
	\renewcommand{\tabcolsep}{0.15cm}
	\vspace{0.1cm}
	\begin{tabular}{cc|c|c|c|c}
		\multicolumn{2}{c}{} & \multicolumn{4}{c}{} \\
		\hline
		\multicolumn{2}{c|}{}  & QRPA-FRDM \cite{Mol03} & LSSM \cite{Zhi13} &  CQRPA-DF3\cite{Bor06}&  pn-QRPA (N) \\
		\hline Nucl.     & A     & $P_{n}$& $P_{n}$ &$P_{n}$  & $P_{n}$\\
		
		\hline
		Tb        & 191    & --    & --   & 72.7 & 100  \\
		Dy        & 192    & 99.3  & 66.0 & 4.76 & 55.0 \\
		Ho        & 193    & 96.9  & 91.8 & 66.7 & 55.0  \\
		Er        & 194    & 96.6  & 13.3 & 26.5 & 50.5  \\
		Tm        & 195    & 27.2  & 80.3 & 47.3 & 0.55  \\
		Yb        & 196    & 4.08  & 3.40 & 0.34 & 0.55  \\
		\hline
	\end{tabular}
\end{table*}


    \end{document}